\documentclass[12pt,authoryear,3p]{elsarticle_modifie}
\usepackage[T1]{fontenc}
\usepackage[utf8]{inputenc}
\usepackage{color}
\usepackage{array}
\usepackage{longtable}
\usepackage{float}
\usepackage{booktabs}
\usepackage{url}
\usepackage{multirow}
\usepackage{tipa}
\usepackage{amsmath}
\usepackage{graphicx}
\usepackage{subscript}
\usepackage{microtype}
\usepackage[unicode=true,
 bookmarks=true,bookmarksnumbered=true,bookmarksopen=true,bookmarksopenlevel=1,
 breaklinks=false,pdfborder={0 0 1},backref=section,colorlinks=true]
 {hyperref}
\hypersetup{
 linktocpage=true,linkcolor=blue,breaklinks,citecolor=green}

\makeatletter

\newcommand{\lyxmathsym}[1]{\ifmmode\begingroup\def\b@ld{bold}
  \text{\ifx\math@version\b@ld\bfseries\fi#1}\endgroup\else#1\fi}

\providecommand{\tabularnewline}{\\}

\usepackage{color}
\usepackage{textcomp}
\usepackage{lineno}
\usepackage{placeins}

\usepackage{extsizes}
\usepackage{array}
\usepackage{url}
\usepackage{amsfonts}
\usepackage{nicefrac}
\usepackage{lipsum}
\usepackage{tabularx}
\usepackage{pslatex}
\renewcommand*{\appendixname}{Annex }

\usepackage{xcolor}
\def\ps@pprintTitle{
 \let\@oddhead\@empty
 \let\@evenhead\@empty
 \def\@oddfoot{}
 \let\@evenfoot\@oddfoot}
\usepackage[nonumberlist,hyperfirst=false]{glossaries}
\newacronym{gpe}{GPE}{Grand Paris Express}
\newacronym{oc}{OC}{Occupational category}
\newacronym{cba}{CBA}{Cost-Benefit Analysis}
\newacronym{iau}{IAU}{Institut d'Aménagement et d'Urbanisme}
\newacronym{tod}{TOD}{Transit-Oriented Development}
\newacronym{rer}{RER}{Réseau Express Régional}
\newacronym{us}{US}{United States}

\usepackage{chngcntr}

 \date{\today}

\@ifundefined{showcaptionsetup}{}{%
 \PassOptionsToPackage{caption=false}{subfig}}
\usepackage{subfig}
\makeatother

\begin{document}
\title{Testing the monocentric standard urban model in a global sample of
cities}
\author[cired,tub]{Charlotte Liotta\corref{cor1}}
\ead{liotta@centre-cired.fr}
\author[cired]{Vincent Viguié}
\author[cired]{Quentin Lepetit}
\cortext[cor1]{Corresponding author}
\address[cired]{CIRED (Ecole des Ponts ParisTech), Site du Jardin Tropical, 45bis,
Av de la Belle Gabrielle, F-94736 Nogent-sur-Marne, France}
\address[tub]{TU Berlin, Straße des 17. Juni 135, D-10623 Berlin, Germany}
\begin{abstract}
Using a unique dataset containing gridded data on population densities,
rents, housing sizes, and transportation in 192 cities worldwide,
we investigate the empirical relevance of the monocentric standard
urban model (SUM). Overall, the SUM seems surprisingly capable of
capturing the inner structure of cities, both in developed and developing
countries. As expected, cities spread out when they are richer, more
populated, and when transportation or farmland is cheaper. Respectively
100\% and 87\% of the cities exhibit the expected negative density
and rent gradients: on average, a 1\% decrease in income net of transportation
costs leads to a 21\% decrease in densities and a 3\% decrease in
rents per m\textsuperscript{2}. We also investigate the heterogeneity
between cities of different characteristics in terms of monocentricity,
informality, and amenities.
\end{abstract}
\maketitle
Accepted manuscript in Regional Science and Urban economics on August
11th, 2022.

\section{Introduction}

Understanding urbanization patterns and modeling urban dynamics have
been a subject of research for at least a century \citep{duranton_chapter_2015}.
These questions have recently been brought back into the spotlight
in the context of current environmental crises, as urban sprawl is
a key driver of both climate change \citep{ipcc_chapter_2014} and
biodiversity losses \citep{ipbes_global_2019,seto_global_2012}.

A well-established framework to tackle these issues is the Standard
Urban Model (SUM), widely used in theoretical and applied papers \citep{brueckner_urban_2001}.
The SUM, initially developed by \citet{alonso_location_1964}, \citet{muth_cities_1969},
and \citet{mills_aggregative_1967}, describes the relationship between
land use, land value, and transportation costs in cities, and relies
on the following mechanisms. Households trade off between housing
size and transportation costs to employment centers, which, in a monocentric
city, results in high bid-rents in city centers, where transportation
costs are low, and low bid-rents in peripheries. Under the assumption
of competitive land markets, private developers will provide a high
housing supply in city centers and a low housing supply in peripheries,
as inelastically supplied land and diminishing returns to factor inputs
imply that taller buildings require higher floor space rents \citep{ahlfeldt_economics_2022}.
From these mechanisms, aggregated results can be derived at the city
level. \citet{wheaton_comparative_1974} showed that, under the assumptions
of monocentricity and monetary commuting costs, large populations
and high incomes lead to larger urbanized areas, while high transportation
costs and high agricultural land prices lead to smaller urban areas.

An extensive literature investigates the validity of the SUM and empirically
examines its main conclusions in terms of population densities, rents,
and urbanized areas, with studies on density gradients starting even
before the theory was formulated \citep{clark_urban_1951}. However,
this literature still suffers from some limitations. First, few studies
simultaneously investigate the model's predictions in terms of urbanized
areas, population densities, and rent gradients in relation to transportation
costs within cities. Indeed, while the literature on density gradients
in cities is well developed (see, for instance, the review by \citealp{mcdonald_econometric_1989}),
the literature on dwelling rents or prices is less developed due to
data limitations, and most studies focus on one or two cities, preventing
comparison between cities or generalization of results \citep{duranton_chapter_2015,mcmillen_testing_2006,mcmillen_issues_2010}.
In addition, most of the existing studies investigate developed countries.
Exceptions include \citet{deng_growth_2008}, \citet{ke_determinants_2009},
\citet{angel_dimensions_2011}, and \citet{jedwab_comparing_2021}
for urbanized area sizes. Exceptions regarding density gradients are
more numerous as the literature is larger but include only three large-scale
studies comparing a large number of cities \citep{bertaud_spatial_2014,mills_comparison_1980,ingram_spatial_1981}
and a few single case studies, mainly in China \citep{chen_new_2010,feng_spatial_2009,wang_modelling_1999}.

However, the recent increase in data availability offers an opportunity
to deepen our understanding of urban forms and city structures globally.
Satellite data can be used to compare urban shapes: for instance,
\citet{harari_cities_2020} uses such data to demonstrate the impact
of city compactness on accessibility and welfare. New datasets, such
as night lights \citep{henderson_measuring_2012} and Twitter data
\citep{indaco_twitter_2020} can also be used as proxies for economic
activity at a disaggregated, e.g. city, level. Regarding transportation,
big datasets, for instance from Google Maps, allow detailed analysis
of mobility and congestion for large samples of cities \citep{akbar_mobility_2018,hanna_citywide_2017}.

Using spatially-explicit data on population densities, rents, dwelling
sizes, land cover, and transport time in 192 cities across all continents,
this paper empirically assesses the ability of the SUM to represent
cities throughout the world. We show that the SUM captures urban structures
surprisingly well, both in developed and developing countries, with
most cities exhibiting the expected negative rent and density gradients:
on average, a 1\% decrease in income net of transportation costs leads
to a 21\% decrease in densities and a 3\% decrease in rents per m\textsuperscript{2},
and moving 1km farther from the city center decreases rents by 1.4\%
and population density by 8.5\%. We structurally estimate that the
median capital elasticity of housing production over the sample of
cities is 0.87, and that the median percentage of housing expenses
in households' income net of transportation costs is 33\%. Our sample
also allows us to investigate heterogeneity between cities: we show
that polycentricity, informality, and local amenities weaken the relationship
between rents or densities and incomes net of transportation costs,
flattening density and rent gradients in cities. Aggregated predictions
of the model in terms of urbanized areas are also verified in our
sample: cities spread out more when they are richer, more populated,
and when transportation or famland rents are less costly.

The remainder of the paper is organized as follows. Section 2 introduces
the monocentric SUM and the extant evidence regarding its main predictions.
Section 3 presents the data. Section 4 empirically tests the three
main relationships derived from the model, and section 5 discusses
the results.

\section{Introducing the monocentric city model}

\subsection{Theory\label{subsec:Theory}}

This section describes the SUM of urban economics and derives expressions
for three endogenous variables (rents within cities, population densities,
and total urbanized area of each city) as a function of a set of exogenous
variables (population, farmland rents, income, per-distance transport
cost). We analyse a sample of cities indexed by \textit{c}, with locations
within cities indexed by \textit{i}, in a closed-city framework assuming
exogenous populations. 

In the first step, we derive dwellings' bid-rents per m\textsuperscript{2}
at location\textit{ i} in city \textit{c}, denoted $R_{i,c}$, as
a function of transportation costs to the city center (where we assume
for simplicity that all jobs are located) $T_{i,c}$ using the spatial
equilibrium condition. We assume that households seek to maximise
their Cobb-Douglas utilities:

\begin{equation}
max\hspace{3mm}U(z_{i,c},\,q_{i,c})=z_{i,c}^{\alpha_{c}}\,q_{i,c}^{\beta_{c}}\hspace{3cm}s.t.\hspace{3mm}z_{i,c}+q_{i,c}R_{i,c}+T_{i,c}\leq Y_{c}\label{eq:max-1}
\end{equation}
with $q_{i,c}$ the size of the dwellings in pixel \textit{i} and
city \textit{c}, $Y_{c}$ the income in city \textit{c} (assumed as
constant over the city), $z_{i,c}$ the consumption of a composite
good (whose price is normalized to 1), and $\alpha_{c}$ and $\beta_{c}$
the Cobb-Douglas parameters (such that $\alpha_{c}+\beta_{c}=1$).

From equation \ref{eq:max-1}, we find that: 
\begin{equation}
\begin{array}{l}
q_{i,c}=\beta_{c}(Y_{c}-T_{i,c})/R_{i,c}\\
z_{i,c}=\alpha_{c}(Y_{c}-T_{i,c})
\end{array}\label{eq:dwelling_size}
\end{equation}
Using the spatial equilibrium condition and writing $u_{c}$ the uniform
utility at equilibrium, we find: 
\begin{equation}
R_{i,c}=R_{0,c}\,\frac{(Y_{c}-T_{i,c})^{1/\beta_{c}}}{Y_{c}^{1/\beta_{c}}}\hspace{1cm}\text{with}\hspace{1cm}R_{0,c}=(\frac{\alpha_{c}^{\alpha_{c}}\,\beta_{c}^{\beta_{c}}\,Y_{c}}{u_{c}})^{1/\beta_{c}}\label{eq:rents}
\end{equation}

In the second step, we derive population density as a function of
transportation costs. We assume that absentee private developers produce
housing from capital $K_{i,c}$ and urbanizable land $L_{i,c}$ with
the following Cobb-Douglas production function, with the parameters
$A_{c}$, $a_{c}$ and $b_{c}$ such that $a_{c}+b_{c}=1$ : 
\begin{equation}
H(K_{i,c},\,L_{i,c})=A_{c}\,L_{i,c}^{a_{c}}\,K_{i,c}^{b_{c}}\label{eq:housing_production_function}
\end{equation}
With $k_{i,c}=K_{i,c}/L_{i,c}$ (capital intensity per urbanizable
land surface area) and $h=H/L_{i,c}$ (housing density, i.e. number
of m2 built upon per m2 of available ground), equation \ref{eq:housing_production_function}
can be rewritten: 
\begin{equation}
h(k_{i,c})=A_{c}k_{i,c}^{b_{c}}
\end{equation}
Developers seek to maximize their profit per land surface area $\pi_{i,c}=R_{i,c}h(k_{i,c})-\rho k_{i,c}$
considering a capital cost $\rho$. Thus, housing supply is written:
\begin{equation}
H_{i,c}=A_{c}^{1/a_{c}}(b_{c}R_{i,c}/\rho)^{b_{c}/a_{c}}L_{i,c}
\end{equation}
and total population per pixel $N_{i,c}$ is written:
\begin{equation}
N_{i,c}=H_{i,c}/q_{i,c}=A_{c}^{1/a_{c}}(b_{c}R_{i,c}/\rho)^{b_{c}/a_{c}}L_{i,c}/q_{i,c}\label{eq:-1}
\end{equation}

Replacing $q_{i,c}$ by its expression in equation \ref{eq:dwelling_size}:
\begin{equation}
N_{i,c}=A_{c}^{1/a_{c}}(b_{c}/\rho)^{b_{c}/a_{c}}L_{i,c}R_{i,c}^{1/a_{c}}\frac{1}{\beta_{c}(Y_{c}-T_{i,c})}\label{eq:eq_density_distance-1}
\end{equation}

The population density per pixel $n_{i,c}=N_{i,c}/L_{i,c}$ can be
expressed as:

\begin{equation}
n_{i,c}=A_{c}^{1/a_{c}}(b_{c}/\rho)^{b_{c}/a_{c}}R_{i,c}^{1/a_{c}}\frac{1}{\beta_{c}(Y_{c}-T_{i,c})}\label{eq:eq_density_distance-1-1}
\end{equation}

Finally, replacing $R_{i,c}$ by its expression in equation \ref{eq:rents}: 

\begin{equation}
n_{i,c}=A_{c}^{1/a_{c}}(\frac{b_{c}}{\rho})^{b_{c}/a_{c}}(\frac{\alpha_{c}^{\alpha_{c}}}{u})^{1/(\beta_{c}a_{c})}\beta_{c}^{b_{c}/a_{c}}(Y_{c}-T_{i,c})^{(1-\beta_{c}a_{c})/\beta_{c}a_{c}}\label{eq:eq_density_distance-1-1-1}
\end{equation}

In the third step, we want to derive the total urban area as a function
of city characteristics. We assume for simplicity that the city is
circular, i.e. that the amount of urbanizable land per pixel is constant,
with $L_{i,c}=1$, and that transportation cost can be broken down
as $T_{i,c}=d_{i,c}t_{c}$ with $d_{i,c}$ the distance from pixel\textit{
i} to the city center of city \textit{c} and $t_{c}$ the constant
per unit transportation cost.

We assume the following equilibrium conditions: 
\begin{itemize}
\item from equation \ref{eq:rents}, we find that bid-rents decrease from
city centers to peripheries. We assume that the bid-rent $R_{i,c}$
intersects with farmland rent $\overline{R}_{c}$ at the fringe of
the city $\overline{d}_{c}$ and at a transport cost $\overline{T}_{c}=\overline{d}_{c}t_{c}$,
such that: 
\end{itemize}
\begin{equation}
\overline{R}_{c}=R_{0,c}\,(1-\frac{\overline{T}_{c}}{Y_{c}})^{1/\beta_{c}}=R_{0,c}\,(1-\frac{t_{c}\overline{d}_{c}}{Y_{c}})^{1/\beta_{c}}\label{eq:agri_rent-1}
\end{equation}

\begin{itemize}
\item in a closed-city model, all the population of the city $N^{*}$ lives
within the city, i.e.
\end{itemize}
\begin{equation}
N_{c}^{*}=\int{}^{i}n_{i,c}di
\end{equation}

\citet[pp. 830-836 and 840-844]{brueckner_structure_1987} shows that,
under the following hypotheses:
\begin{itemize}
\item preference of households for housing and composite good is represented
by a quasi-concave utility function, as in equation \ref{eq:max-1};
\item housing supply is represented by a concave production function with
constant returns, as in equation \ref{eq:housing_production_function};
\end{itemize}
we can derive the following relationships:

\begin{equation}
\frac{\text{\ensuremath{\partial}}\overline{d}_{c}}{\partial N_{c}^{*}}>0,\frac{\text{\ensuremath{\partial}}\overline{d}_{c}}{\text{\ensuremath{\partial}}Y_{c}}>0,\frac{\text{\ensuremath{\partial}}\overline{d}_{c}}{\text{\ensuremath{\partial}}t_{c}}<0,\frac{\text{\ensuremath{\partial}}\overline{d}_{c}}{\text{\ensuremath{\partial}}\overline{R}_{c}}<0,\label{eq:urban_area_eq}
\end{equation}

Therefore, urban areas are expected to decrease with transportation
cost per unit of distance and farmland rent and increase with total
population and income.

\subsection{Extant evidence\label{subsec:Extant-evidence}}

This section reviews the extant literature and evidence regarding
the three relationships derived from the SUM in the previous subsection:
bid-rents expressed as a function of transportation costs (equation
\ref{eq:rents}), population density as a function of transportation
costs (equation \ref{eq:eq_density_distance-1-1-1}), and urbanized
area as a function of the total population, the average income, transportation
costs, and farmland rents (equation \ref{eq:urban_area_eq}).

\subsubsection{Rent gradients \label{subsec:ev-Rent-gradients}}

Studies beginning as far back as the 1970s have measured rent or land
price gradients within cities. Most of them have found evidence of
negative rent or land price gradients from city centers to the suburbs,
mainly in the US \citep{coulson_really_1991} and the UK \citep{cheshire_price_1995},
and more recently in Berlin \citep{ahlfeldt_if_2011}.

Nevertheless, flat or inverted rent or land price gradients have been
measured in some cities. For instance, \citet{yinger_estimating_1979}
found the expected negative rent gradients in Madison, in the USA,
but not in the more polycentric setting of St Louis. Working on Chicago,
\citet{mcdonald_land_1979} found a negative land price gradient close
to the Central Business District (CBD) but an inverted land price
gradient farther away, which they explained by polycentrism, racial
segregation, or disamenities such as air pollution or congestion in
the city center. Also in Chicago, \citet{mcmillen_one_1996} showed
that the distance to the CBD could satisfactorily explain land values
in the city between 1836 and 1928, but that it was no longer significant
after 1960, as the development of O'Hare Airport made the city more
polycentric.

Beyond accessibility, other variables have been shown to impact rents
or land values, modifying the gradient. \citet{cheshire_price_1995}
found the expected negative land price gradient in Reading and Darlington,
in the UK, but they use a fully hedonic model and show that, in addition
to accessibility, location-specific and dwelling-specific amenities
play a key role in determining rents. More recently, \citet{ahlfeldt_if_2011}
showed that the monocentric SUM performs satisfactorily if we account
for structural and neighborhood characteristics such as distances
to the nearest school or green space. He also showed that gravity
employment accessibility measures perform better than the monocentric
SUM in the polycentric setting of Berlin.

To summarize, despite the early literature on rent and land value
gradients, the evidence for the negative relationship between rent
and distance to CBD is mixed. Heterogeneity in rent gradients between
cities can be explained by polycentricity \citep{yinger_estimating_1979,mcmillen_one_1996}
and location-based and dwelling-based amenities \citep{cheshire_price_1995}.
A more systematic measure of rent gradients in various cities and
an evaluation of the explanatory power of the hypotheses that have
been proposed in the literature would usefully complement the existing
studies.

\subsubsection{Density gradients}

The literature concerning population density gradients, summarized
in \citet{mcdonald_econometric_1989} and \citet{duranton_chapter_2015},
is wider, since data on population densities are easier to obtain
than data on rents or land prices. An initial line of research, starting
even before the formulation of the theory \citep{clark_urban_1951},
sought to empirically find regularities in population densities. Many
studies support the negative exponential function \citep{berry_urban_1963,weiss_distribution_1961},
theoretically grounded in \citet{muth_spatial_1961}. However, other
functions have been proposed in the literature, including the addition
of a quadratic component to the negative exponential function to describe
the population crater near the city center \citep{newling_spatial_1969,latham_population_1970},
the inverse square function \citep{stewart_physics_1958,batty_form_1992},
and the Gaussian function \citep{ingram_concept_1971,guy_assessment_1983}.
A full review of these functional forms can be found in \citet{martori_urban_2002},
and \citet{qiang_shapes_2020} performed a recent assessment of their
explanatory power on American cities. Overall, all these studies agree
that population densities decrease from the city center to the periphery,
potentially attenuated by the city center population density crater,
due to the concentration of economic activities.

A second line of research has compared density gradients, either in
the same city across time or between cities. Two main results emerge
from this literature. First, there is evidence of a flattening of
density gradients over time in Latin America \citep{ingram_spatial_1981}
and the US \citep{mills_urban_1970,macauley_estimation_1985,kim_changes_2007},
which might be due to an increase in incomes and a decrease in transportation
costs \citep{mills_urban_1970}. Second, this gradient depends on
the characteristics of the cities. More populous cities tend to have
flatter gradients \citep{muth_cities_1969,guest_urban_1973,edmonston_population_1976,glickman_modeling_1978,johnson_urban_1980,mills_comparison_1980,alperovich_determinants_1983,bertaud_spatial_2014}:
one explanation is that employment subcenters are more likely to emerge
in large areas. Older urban areas tend to have steeper gradients,
possibly because of the inertia in urban shapes: cities that developed
when transportation costs were high are more likely to be compact
(see the studies by \citealp{glickman_modeling_1978}, in Japan and
\citealp{johnson_urban_1980}, in the US). Overall, income has been
found to have a weak negative impact on the steepness of the gradient
\citep{mills_comparison_1980}. Comparing developed and developing
countries, \citet{mills_comparison_1980} found that developed countries
have flatter density functions than developing countries, as they
have higher incomes and better urban transportation systems. Working
on a diverse sample of 57 cities, \citet{bertaud_spatial_2014} found
that cities subject to stringent regulations, such as Brasilia, or
complex geographies, such as coastal cities, exhibit a flatter gradient.

Recently, the increase in data availability has allowed studies to
be performed on large samples of cities (see for instance \citealp{lemoy_evidence_2020},
working on 300 European cities) or with better proxies for transportation
costs. \citet{qiang_shapes_2020}, for instance, use transportation
time data from Google Maps instead of distances to city centers. These
data could allow empirical assessment, using larger samples and with
increased accuracy, of the extant results relating to density gradient
heterogeneity between cities of different sizes or degrees of monocentricity
found in the literature.

\subsubsection{Urban areas}

Finally, the following relationship can be derived from the SUM with
respect to urban areas: their sizes should depend positively on population
and income and negatively on transportation costs and farmland rents
(equation \ref{eq:urban_area_eq}). Many papers have empirically verified
these results in the US \citep{brueckner_economics_1983,mcgrath_more_2005,wassmer_influence_2006,spivey_millsmuth_2008,paulsen_yet_2012},
finding that population, income, and farmland rents have the expected
impact on urban areas. An important exception is \citet{spivey_millsmuth_2008},
estimating a negative income elasticity, which she explains by the
fact that the increase in incomes increases the demand for more affordable
housing farther from the city center, but also raises the opportunity
cost of time, the second effect outweighing the first. However, these
studies struggled to find proxies for transportation costs, leading
to mixed results. \citet{spivey_millsmuth_2008} uses the fraction
of households owning a car, which is not robust in her main specification,
and the Texas Transportation Institute’s “time travel index” and “congestion
cost index,” with which she finds mixed evidence. \citet{paulsen_yet_2012}
does not include any commuting cost variables: the author considered
the census average travel time to work, but finally dropped this variable.

Fewer studies exist in Europe. \citet{oueslati_determinants_2015}
found conclusive results, with all variables being significant and
of the expected sign. In particular, the authors found a high coefficient
for farmland rents, which reflects the fact that in Europe, agriculture
at the urban fringe is often highly intensive, offering relatively
high yields and profits. In Germany, \citet{schmidt_does_2020} found
a particularly low coefficient for income, which they explain by the
fact that increased demand for housing is not realized due to growth
management policies and restrictive land-use regulations. Their proxies
for transportation costs were found to be unreliable, leading to insignificant
estimates, or to estimates whose sign is opposite to that which is
expected. They also found differences between east and west Germany
and between growing and shrinking regions.

Few studies focused on developing countries. The existing studies
on China \citep{deng_growth_2008,ke_determinants_2009} found that
income growth played a significant role in determining urban sprawl.
Recently, \citet{jedwab_comparing_2021} showed that cities in developing
countries grow differently to those in developed countries, i.e. by
crowding ``in'' instead of sprawling or growing ``up''.

At a global level, analyzing both total urban land cover in 206 countries
and urban land cover in 3646 large urban areas worldwide, \citet{angel_dimensions_2011}
found significant estimates of the expected sign, except for transportation
costs. However, the authors do not analyze the heterogeneity of their
results by world regions or development level.

\section{Data \label{section:data}}

\subsection{Spatially-explicit data}

Our main dataset, described in detail in \citet{lepetit_gridded_2022},
is a spatially-explicit dataset including population densities, rents,
dwelling sizes, transportation times, and land cover in 192 cities
worldwide. This dataset's originality lies in the provision of data
on cities' internal structures at a 1km\textsuperscript{2} resolution,
in particular on rents and transportation times, while covering a
large and diverse sample of 192 cities on five continents. By combining
data on density, real estate, transportation, and land cover, this
dataset allows us to run an integrated analysis of the validity of
the SUM in the sample. The main data collection steps are summarized
below.

\begin{figure}[H]
\centering\includegraphics[width=1\columnwidth]{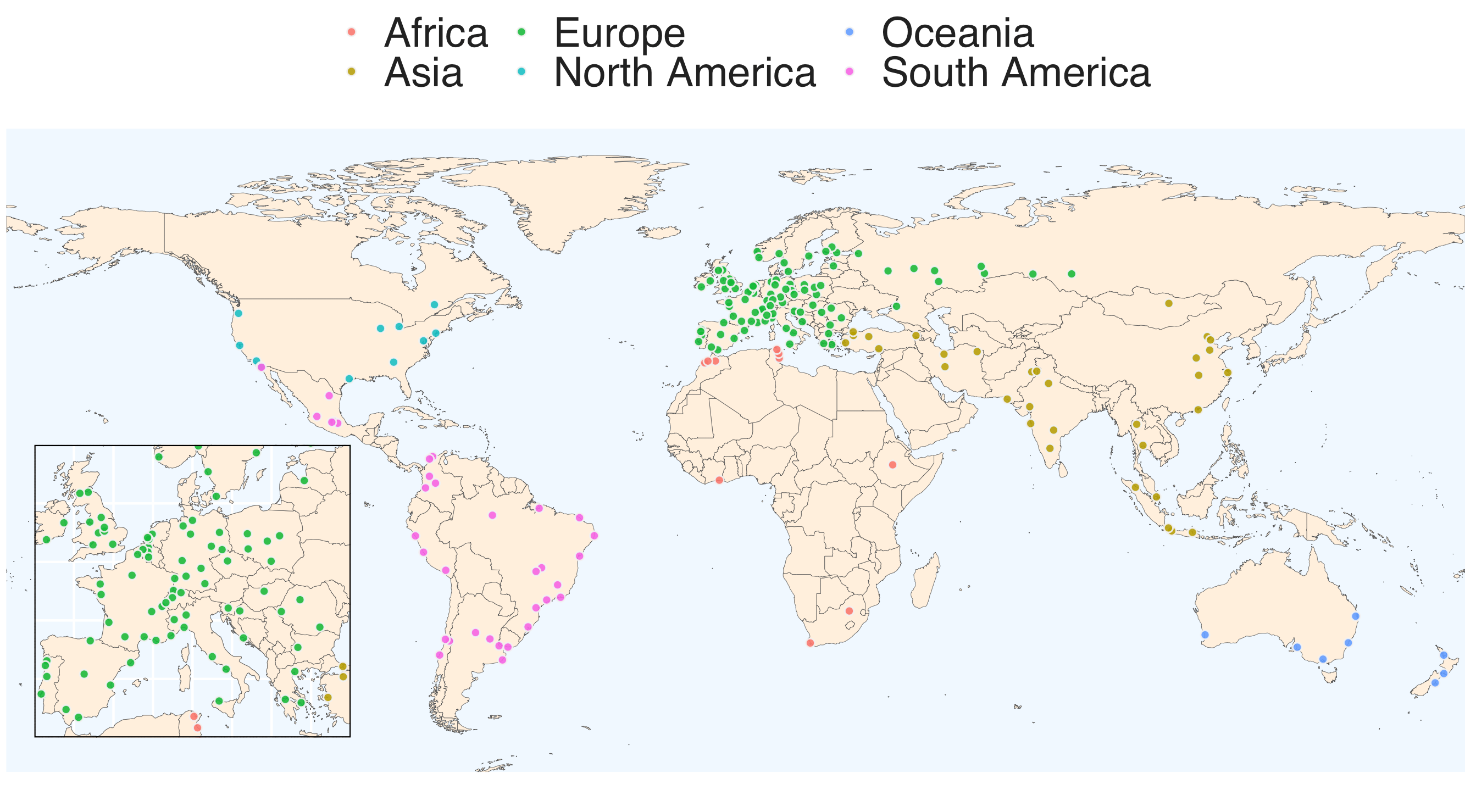}
\caption{Sample of cities for which data were collected.}
\label{fig:data}
\end{figure}

The cities were chosen to obtain wide geographical coverage, and represent
different cultural and historical backgrounds. However, data availability,
in particular for real estate, has been a limiting criterion. The
cities selected are medium to large cities: they all have more than
300,000 inhabitants, and together represent 800 million people, i.e.
19\% of the global urban population, or 34\% of the people living
in cities of more than 300,000 inhabitants (Figure \ref{fig:data}).

First, we divided each city into a georeferenced 1 km\textsuperscript{2}-
resolution grid encompassing the whole urban area. Then, we combined
data from different sources and aggregated them on these grids. For
land use, we used the European Space Agency land cover data \citep{esa_land_2017},
available worldwide at a 300m spatial resolution on an annual basis
from 1992 to 2015. We reclassified areas as “urbanizable\textquotedbl{}
(e.g., farmlands) or “non-urbanizable” (e.g., water bodies) (\ref{appendix:land_cover}).
For population density, we used the European Commission GHSL data
available for 1975, 1990, 2000, and 2015 at a 250m resolution worldwide
\citep{schiavina_ghs_2019}.

Data on dwelling rents per m\textsuperscript{2} and dwelling sizes
were most difficult to obtain, as no global database exists. We obtained
them by web-scraping real estate websites between 2017 and 2020. The
websites were selected to meet four criteria: the website must have
nationwide coverage to ensure consistent results in a given country,
it must geolocalize the dwellings, have values for both rent or sale
prices and dwelling size, and be written in the local language and
give prices in local currency to limit real estate ads targeting expatriates.
We aggregated the data at the grid cell level by taking mean rents
per square meter. We computed two variables to assess rent data quality
(Table \ref{tab:Real-estate-data-quality}): first, \textquotedbl Real
estate data market cover\textquotedbl{} which corresponds to the average
number of real estate ads that were scraped per grid cell in the city\footnote{This is given by (population/number of ads).}
and second, \textquotedbl Real estate data spatial cover\textquotedbl ,
the percentage of grid cells that have density data and for which
we also have data on rents\footnote{This is given by (number of pixels with real estate data)/(number
of pixels with population > 0).}. In 109 cities out of 192, the spatial data cover of rent ads is
above 10\%, and in 153 cities out of 192, the spatial data cover is
above 5\%. In 95 cities out of 192, we scraped more than 1 rent ad
per 1,000 inhabitants, and in 174 cities out of 192, we scraped more
than 1 rent ad per 10,000 inhabitants.

\begin{table}[H]
\begin{tabular}{cccccccc}
 & {\small{}Mean} & {\small{}Min.} & {\small{}Q1} & {\small{}Med.} & {\small{}Q3} & {\small{}Max.} & {\small{}Nb. of obs.}\tabularnewline
\hline 
{\small{}Market cover} & {\small{}14168} & {\small{}98} & {\small{}550} & {\small{}1020} & {\small{}2281} & {\small{}810960} & {\small{}192}\tabularnewline
{\small{}Spatial cover} & {\small{}11.81\%} & {\small{}0.07\%} & {\small{}5.79\%} & {\small{}10.90\%} & {\small{}15.41\%} & {\small{}59.14\%} & {\small{}192}\tabularnewline
\hline 
\end{tabular}\caption{Real estate data quality.\protect \\
\textit{For 50\% of the cities, we scraped more than 1 real estate
ad per 1,020 inhabitants, and we scraped real estate ads for more
than 10.9\% of the inhabited pixels.} \label{tab:Real-estate-data-quality}}
\end{table}

Transport distances and durations from each grid cell to the city
center have been estimated using Google Maps and Baidu Maps APIs (Application
Programming Interfaces). Different definitions of city centers exist
in the urban economics literature. Most rely on job density data,
which are unfortunately not consistently available for the cities
in the database. Therefore, this study defined city centers on the
basis of a compromise between five qualitative criteria: the geographical
center of the data, the historical center of the city, the location
of the public transport hub, the official central business district,
and the city hall location. When available, both driving and public
transport data were collected for typical afternoon rush hours\footnote{We have identified rush hours for each city by extracting a sample
of transportation data at 16h, 16h30, 17h, 17h30, 18h, 18h30, 19h,
19h30, and 20h, by comparing the average delay due to congestion,
and by defining rush hours as the moment when the average delay is
highest.}. It was not possible to collect transport data from each grid cell,
so 10\% of all cells were collected\footnote{With a method close to \citet{saiz_physical_2021}, we defined a star
shape with 8 branches, centered on the city center, and we collected
data from the grid cells at a regular distance along each star branch
so that the number of grid cells for which data were collected equals
10\% of the total number of grid cells.} and then performed bivariate interpolation using the R package AKIMA,
allowing us to approximate transportation times for the whole grid.

\subsection{Aggregated data at the city level}

To complement the spatially-explicit dataset, we use data relating
to characteristics of the whole cities, detailed below.

City populations are derived from the spatially explicit dataset and
correspond to the sum of the densities over the grid. Similarly, urbanized
areas correspond to the sum of the urbanized area data over the grid.
We are aware that this definition of city limits is likely to overestimate
population and urbanized area, but it allows the extent of \textquotedbl urban
area\textquotedbl{} to be identical for the two variables and independent
of city administrative or legally-defined boundaries. Indeed, the
extent of functional urban areas may differ substantially from cities'
administrative limits: administrative or legally-defined city boundaries
tend to adapt slowly to rapid changes in population and economic activities
and are highly inconsistent between countries \citep{moreno-monroy_metropolitan_2021}.

Incomes are approximated using World Bank GDP per capita data at the
country level. We also use World bank data on gasoline prices and
\citet{international_energy_agency_iraq_2012} data on fuel efficiency
to compute the cost of travel by car. We estimate farmland rents with
data from the FAO, dividing agricultural GDP by the total agricultural
area in the country.

We measure monocentricity using a qualitative index that shows the
extent to which the geographical center, the historical center, the
public transport hub, the official CBD, and the town hall are located
in the same place. The higher this index, the more monocentric the
city. For instance, in Paris, the historical center (Notre Dame),
the transport hub (Châtelet - Les Halles), and the town hall (Hôtel
de Ville) are very close to each other: only the official CBD (la
Défense) is located further away, at 8km from the city center. This
is a rough proxy compared to the approach used in other studies (see,
for instance, \citealp{ahlfeldt_prime_2020}), but no consistent database
of employment centers is available for all the sampled cities. 

We use data on the percentage of informal housing per country and
Gini indices per country from the World Bank. We have also designed
a categorical variable for the regulatory environment. This variable,
based on \citet{bertaud_spatial_2014}, is equal to 2 for “stringently
planned cities in violation of market principles” (such as Moscow
and Brasilia), to 1 for cities, like Warsaw or Sofia, that have imposed
stringent planning onto a market-oriented base and to 0 for market-oriented
(though still planned) cities. This variable has been taken from \citet{bertaud_spatial_2014}
for the 45 cities in common between their sample and ours. For the
remaining cities, we used our own judgement to allocate this variable:
we looked at whether the Internet-based information available for
each city, relying mainly on Wikipedia, mentions a widespread, stringent
planning approach. The table showing this variable can be found in
\ref{appendix:Regulatory-environmnent}.

\section{Testing the monocentric city model}

\subsection{Rent gradients \label{subsec:Rent-elasticities}}

\subsubsection{Empirical strategy}

This subsection empirically assesses the relationship between rents
per m\textsuperscript{2} and transportation times. Building on equation
\ref{eq:rents}, we run the following regression on the cities in
the database:

\begin{equation}
ln(R_{i,c}){\displaystyle =\,}e_{c}+f_{c}ln(Y_{c}-T_{i,c})+\epsilon_{i,c}\label{eq:empirical_rent}
\end{equation}

We run this regression 191\footnote{In this section, the city of Sfax is excluded from the sample because
of insufficient real estate data quality.} times to individually estimate $e_{c}$ and $f_{c}$ for all the
sampled cities. $e_{c}$ is a constant, and $f_{c}$ can be interpreted
structurally as $1/\beta_{c}$ , drawing on equation \ref{eq:rents}.
Finally, the error term $\epsilon_{i,c}$ accounts for the fact that,
in reality, locations within cities differ in attributes other than
transportation costs, such as neighborhood-based or dwelling-based
amenities \citep{cheshire_price_1995,ahlfeldt_if_2011}. 

Our empirical approach, theoretically grounded, differs from the existing
studies by using a measure of income net of transportation costs instead
of transportation costs alone. In practice, we assume that incomes
are constant within cities so that incomes net of transportation costs
only vary with transportation costs. This approach is equivalent to
the majority of the existing literature, which directly regresses
rents on a proxy of transportation costs, except that it allows us
to interpret parameter $f_{c}$ as $1/\beta_{c}$. However, a monetary
measure of transportation cost is required to compute $Y_{c}-T_{i,c}$:
we thus cannot simply measure transportation costs as the Euclidean
distance to the city center, as in most existing studies.

Like \citet{qiang_shapes_2020}, we use commuting times and distances
from Google Maps data. We go two steps further by considering two
transportation modes and accounting for both opportunity and monetary
costs. We compute transportation costs as follows:
\begin{itemize}
\item We assume that households choose the cheaper transportation mode between
private cars and public transport;
\item For each transportation mode, we derive the opportunity cost of time
from Google Maps' transportation time data assuming that workers value
it at the hourly wage rate;
\item We compute the monetary cost of private cars as the product of the
commuting distance, vehicle efficiency, and fuel prices;
\item We assume that the monetary cost of public transport is a fixed cost
per trip, thus independent of distance: this varies between cities
and has been obtained from various sources.
\end{itemize}
However, this measure of transportation costs might be endogenous.
Indeed, it is based on real-world speeds, which depend on congestion
and infrastructure and might capture negative externalities of transport
infrastructures such as noise or barrier effects. To overcome this
limitation, we use the logarithm of the Euclidean distance to the
city center as an instrumental variable (2SLS approach). Indeed, changes
in distances to the city center are likely to impact transportation
costs, but they have little impact on transportation-infrastructure-related
amenities or disamenities.

\subsubsection{Results}

We start by analyzing the sign of parameter $f_{c}$. Based on the
theory, we expect that rents will increase with income net of transportation
costs, leading to a positive estimate of $f_{c}$. However, the existing
literature has found mixed evidence regarding the relationship between
rents and transportation costs (see subsection \ref{subsec:ev-Rent-gradients}).

\begin{figure}[H]
\includegraphics[scale=0.6]{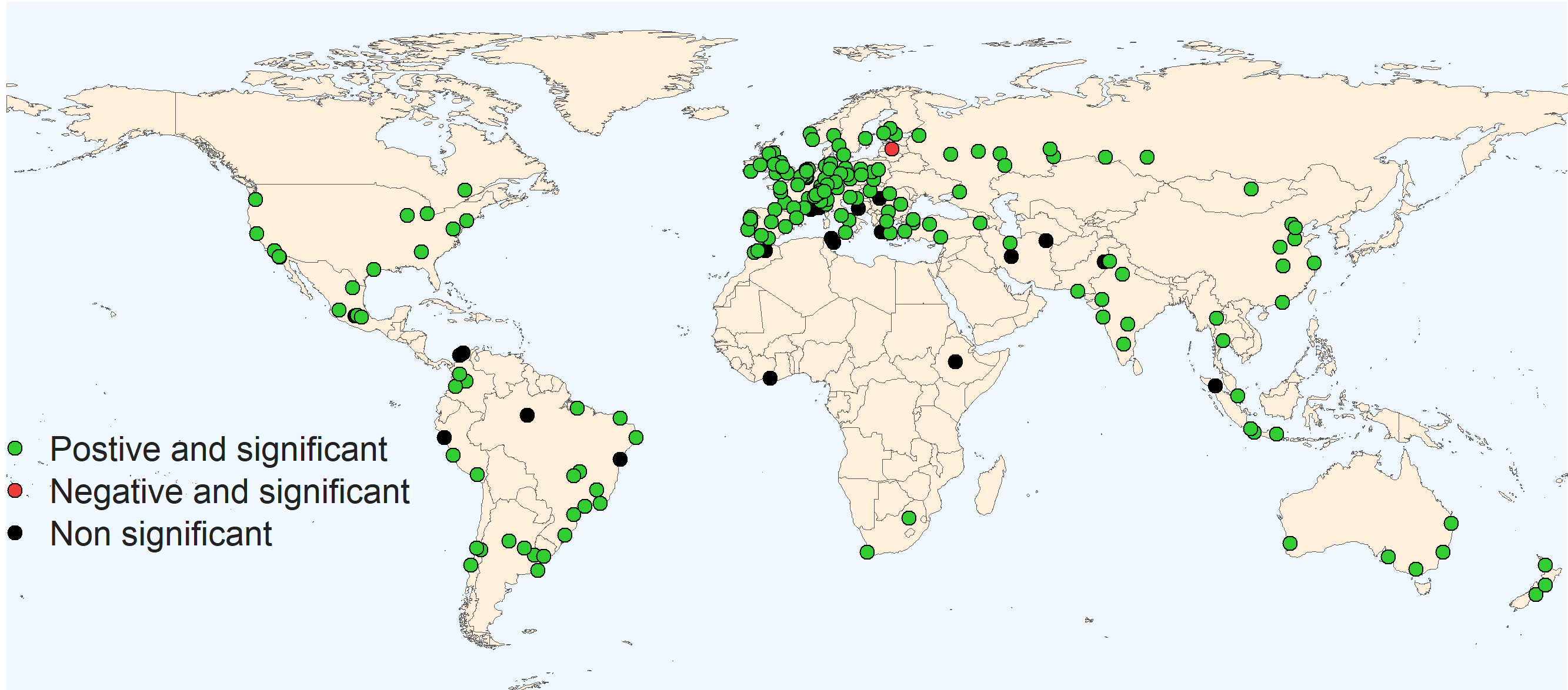}\caption{Map of the direction and significance of parameter $f_{c}$ (rent
elasticity with respect to income net of transportation costs), estimated
with 2SLS. \protect \\
\textit{This parameter is positive and significant in 167 cities,
negative and significant in one city, and non-significant in 23 cities.}\label{fig:sign_rent_grad}}
\end{figure}

Figure \ref{fig:sign_rent_grad} shows, for each city, based on the
2SLS specification, whether the estimate for parameter $f_{c}$ is
positive and significantly different from 0 at the 10\% level, negative
and significant at the 10\% level, or non-significant at the 10\%
level. We find, as expected, that the parameter is positive and significant
at the 10\% level in the vast majority of the cities (167 with 2SLS).
However, we find that it is non-significantly different from 0 in
23 cities with 2SLS, including cities where our data are of low quality
(Isfahan, Mashhad, Fez), cities with a large percentage of informal
housing (Addis Abeba, Abidjan, Manaus), and coastal cities where coastal
amenities are likely to impact rent gradients (Nice, Patras). It is
negative in one city, Riga.

To understand whether the relationship between accessibility and rents
theorized in the SUM is verified, we also compute the R\textsuperscript{2}
of regression (Table \ref{tab:r2_rent}), which can be interpreted
as the percentage of the variance in rents per m\textsuperscript{2}
that is due to the variance in transportation costs. We find that
the mean R\textsuperscript{2} is 0.20 and that R\textsuperscript{2}
can be up to 0.66 in some cities (OLS), supporting the postulate that
rents strongly depend on transportation costs to the city center but
showing significant heterogeneity between cities.

\begin{table}[H]
\centering%
\begin{tabular}{cccccccc}
 & Mean & Min. & Q1 & Med. & Q3 & Max. & Nb. of obs.\tabularnewline
\hline 
OLS & 0.20 & 0.00 & 0.06 & 0.18 & 0.31 & 0.66 & 191\tabularnewline
2SLS & 0.19 & -0.17 & 0.04 & 0.16 & 0.30 & 0.66 & 191\tabularnewline
\hline 
\end{tabular}\caption{Distribution of the R\protect\textsuperscript{2} of equation \ref{eq:empirical_rent}.\label{tab:r2_rent}}
\end{table}

\begin{figure}[H]
\centering

\subfloat[Histogram - distribution of parameter \textit{f}\protect\textsubscript{\textit{c}}\textit{.}]{\centering\includegraphics[scale=0.7]{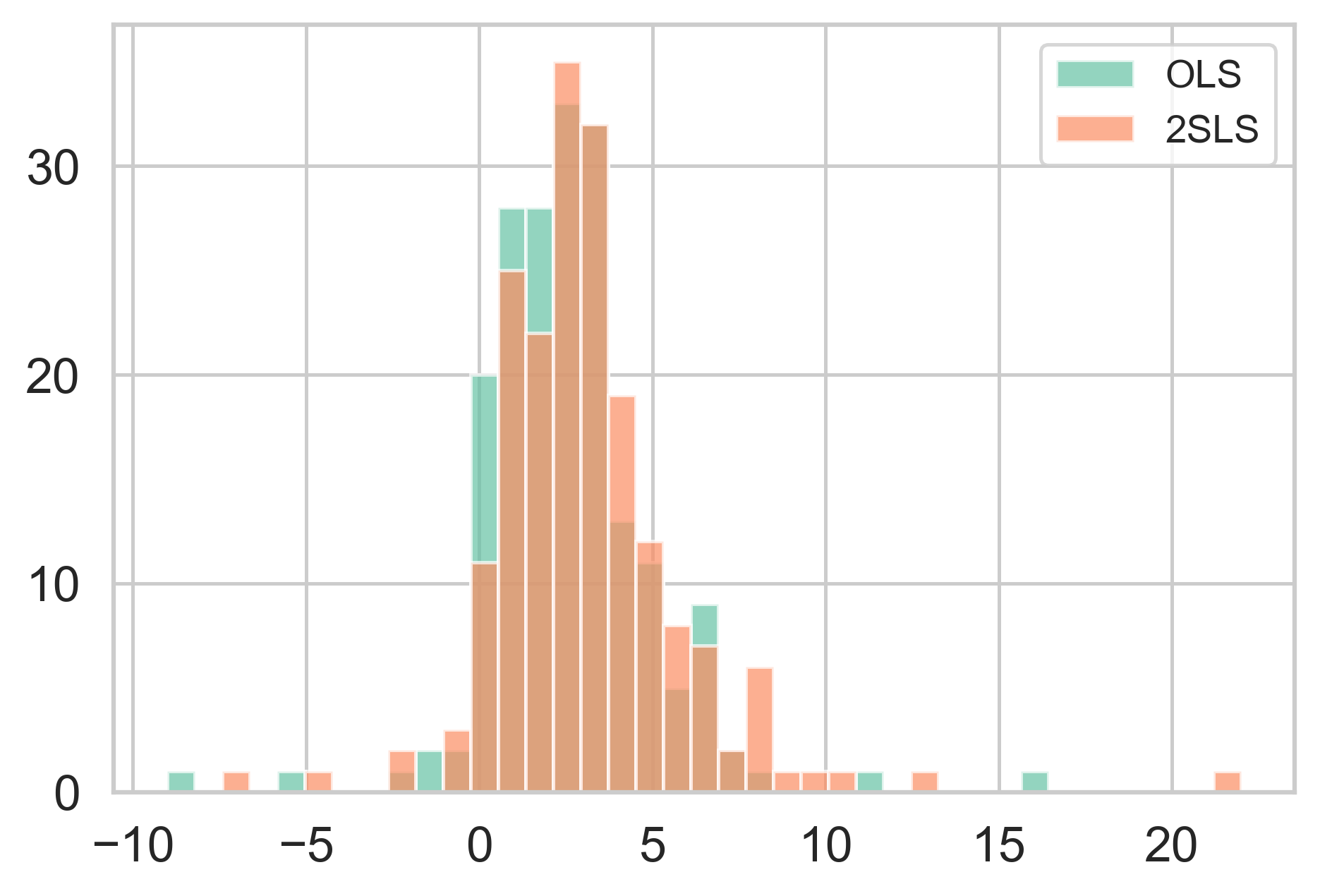}

}

\subfloat[Table - distribution of parameter \textit{f}\protect\textsubscript{\textit{c}}\textit{.} ]{\centering%
\begin{tabular}{cccccccc}
 & Mean & Min. & Q1 & Med. & Q3 & Max. & Nb. of obs.\tabularnewline
\hline 
OLS & 2.59 & -8.83 & 1.10 & 2.33 & 3.51 & 16.14 & 191\tabularnewline
2SLS & 3.10 & -6.78 & 1.50 & 2.71 & 4.13 & 21.78 & 191\tabularnewline
\hline 
\end{tabular}

}\caption{Distribution of the estimates of parameter \textit{f}\protect\textsubscript{\textit{c}}
(rent elasticity with respect to income net of transportation costs)
in the 191 cities.\label{fig:Distrib-rent-grad}}

\end{figure}

Then, we analyze the value of the estimate of $f_{c}$. Figure \ref{fig:Distrib-rent-grad}
shows the distribution of these estimates. Despite a few negative
or very high values (above 8), most estimates are between 0 and 6,
with a median estimate of 2.71 and a mean of 3.10 (2SLS): on average
for our sample of cities, when income net of transportation costs
increases by 1\%, rents increase by 3.1\%. This can be interpreted
as $\beta_{c}$ being around 1/3 in most cities, meaning that households
spend around one-third of their income net of transportation costs
on housing, which is consistent with the data available for the OECD
countries\footnote{OECD affordable housing database, \url{https://www.oecd.org/housing/data/affordable-housing-database/housing-conditions.htm},
accessed on May 17, 2022}.

\subsubsection{Second-step regressions}

This section further examines heterogeneity in the estimates of parameter
\textit{f}\textsubscript{\textit{c}}. We aim to understand why rents
are highly dependent on transportation costs to the city center in
some cities, leading to high rent gradients, while this is not the
case in others, leading to low or even inverted gradients. For that
purpose, we regress our estimates of parameter \textit{f}\textsubscript{\textit{c}}
on a set of city characteristics based on the literature and on the
intuitions derived from Figure \ref{fig:sign_rent_grad} (Table \ref{tab:rent_grad_second_step}).

In the literature, transportation costs to the city center have been
found to be good predictors of rents, but their explanatory power
can be mitigated by polycentrism and local amenities (see section
\ref{subsec:Extant-evidence}). Therefore, in specification (1), we
regress estimates of parameter \textit{f}\textsubscript{\textit{c}}
on monocentricity and on a dummy indicating whether the city is coastal,
since the sea is in general associated with strong locational amenities
that are likely to impact the real estate market. We also control
for rent data quality and for population, as subcenters are more likely
to emerge in more populous cities. We find that the relationship between
income net of transportation costs and rents is stronger in non-coastal
cities. Transportation costs to the city center are also a better
predictor of rents in monocentric cities, but this coefficient is
non-significant. Finally, as expected, we find a stronger relationship
between rents and income net of transportation costs when our real
estate data are of better quality.

In specification (2), we add three variables: the Gini index, the
percentage of informal housing, and the stringency of the regulatory
environment. The first two are based on our observation that the estimate
of \textit{f}\textsubscript{\textit{c}} seems to be non-significant
in some cities with high inequalities and levels of informal housing.
The third is based on \citet{bertaud_spatial_2014}, who observe that
the regulatory environment might affect the city's structure. We find
that the estimate of parameter \textit{f}\textsubscript{\textit{c}}
is lower in cities with a large percentage of informal housing. We
also find higher estimates in cities for which we have good-quality
data. The dummy for coastal cities is no longer significant. The categorical
variable coding for stringent regulations is also non-significant
at the 10\% level: an explanation might be that, even in a stringently
regulated city such as Moscow with a population density pattern that
departs from the monocentric SUM pattern, housing prices may have
initially been random, but over time consumers tend to value accessibility
and housing prices tend to follow the negative density gradient, similarly
to market economies \citep{bertaud_spatial_2014,bertaud_socialist_1997}.

Specification (3) includes continental fixed effects, taking Europe
as the reference. In this specification, the percentage of informal
housing is significant at the 1\% level; we also find that the relationship
between transportation costs and rents is stronger in Asian cities.
Overall, in these three specifications, R\textsuperscript{2} values
vary between 0.073 and 0.227: we are only able to provide limited
explanations of the heterogeneity in rent gradients between cities.
This might be due to unobserved variables: for instance, the only
amenity we account for is that provided by a coastal location, and
we cannot account for other location-based or dwelling-based amenities.
Another explanation might be quality of variables. The \textquotedbl regulatory
environment\textquotedbl{} dummy, for instance, is based in some cases
on the authors' judgement and might lack accuracy. Finally, some of
the explanatory variables might be correlated, such as the quality
of real estate data and the percentage of informal housing.

\begin{table}[H]
\centering%
\begin{tabular}{cccc}
\hline 
 & \multicolumn{3}{c}{\textit{Dependent variable: rent gradients f}\textsubscript{\textit{c}}}\tabularnewline
\cline{2-4} \cline{3-4} \cline{4-4} 
 &  &  & \tabularnewline
 & (1) & (2) & (3)\tabularnewline
\hline 
 &  &  & \tabularnewline
Coastal city & -0.700$^{*}$ & -0.602 & -0.395\tabularnewline
 & (0.408) & (0.408) & (0.409)\tabularnewline
Monocentricity index & 0.196 & -0.318 & 0.567\tabularnewline
 & (0.426) & (0.479) & (0.487)\tabularnewline
log(population) & -0.132 & 0.161 & -0.255\tabularnewline
 & (0.183) & (0.221) & (0.236)\tabularnewline
Market data cover & -0.000 & -0.000 & -0.000$^{**}$\tabularnewline
 & (0.000) & (0.000) & (0.000)\tabularnewline
Spatial data cover & 7.474$^{***}$ & 6.707$^{**}$ & 3.196\tabularnewline
 & (2.625) & (2.657) & (2.593)\tabularnewline
Gini index &  & -0.045 & -0.037\tabularnewline
 &  & (0.030) & (0.044)\tabularnewline
Percentage of informal housing &  & -0.035$^{*}$ & -0.091$^{***}$\tabularnewline
 &  & (0.021) & (0.026)\tabularnewline
Regulatory environment &  & 0.265 & 0.374\tabularnewline
 &  & (0.413) & (0.404)\tabularnewline
Asia &  &  & 4.065$^{***}$\tabularnewline
 &  &  & (0.919)\tabularnewline
Africa &  &  & 0.329\tabularnewline
 &  &  & (1.213)\tabularnewline
Oceania &  &  & -1.275\tabularnewline
 &  &  & (0.994)\tabularnewline
North America &  &  & -0.060\tabularnewline
 &  &  & (0.946)\tabularnewline
South America &  &  & 1.429\tabularnewline
 &  &  & (0.987)\tabularnewline
Constant & 4.035 & 2.955 & 6.576$^{**}$\tabularnewline
 & (2.868) & (3.029) & (3.213)\tabularnewline
\hline &  &  & \tabularnewline
Observations & 191 & 191 & 191\tabularnewline
$R^{2}$ & 0.073 & 0.104 & 0.227\tabularnewline
Adjusted $R^{2}$ & 0.048 & 0.065 & 0.170\tabularnewline
F Statistic & 2.904$^{**}$ & 2.647$^{***}$ & 4.002$^{***}$\tabularnewline
\hline\hline &  &  & \tabularnewline
\textit{Note:} & \multicolumn{3}{c}{$^{*}$p$<$0.1; $^{**}$p$<$0.05; $^{***}$p$<$0.01}\tabularnewline
\end{tabular}\caption{Second-step regression of the estimates of rent gradients  \textit{f}\protect\textsubscript{\textit{c}}
against city characteristics.\label{tab:rent_grad_second_step}}

\end{table}

\subsubsection{Robustness checks}

We run some robustness checks for alternative specifications of equation
\ref{eq:empirical_rent}. In particular, we regress the logarithm
of rents per m\textsuperscript{2} on the Euclidean distance from
the city center, instead of considering the logarithm of the income
net of transportation costs. On average, in the sample of cities,
we find a gradient of -0.014, meaning that going 1km farther from
the city center decreases rents per sqm by 1.4\% (\ref{sec:Robustness-checks},
Figure \ref{fig:distrib_rent_robustness}). We also find heterogeneity
in the estimates of this gradient between cities: the interquartile
range is -0.019 - -0.006, and the results of the second-step regression
show that this gradient is steeper in Asia and South America, and
flatter in cities with a large percentage of informal housing.

We run some additional robustness checks, including regressing the
logarithm of rents per m\textsuperscript{2} on the logarithms of
transportation costs or transportation times instead of regressing
them on the logarithm of incomes nets of transportation costs. Full
details and results are in \ref{sec:Robustness-checks}. Overall,
our results are robust to alternative specifications. Depending on
the robustness check performed, a rent gradient of the expected direction
is found in between 157 and 167 cities. The percentage of informal
housing, the data quality, and the continental fixed-effects are found
to impact the value of the rent gradient in all robustness checks.

\medskip{}
To summarize this subsection, despite the rent gradient having rarely
been studied in the literature, in particular in developing countries,
we found a gradient of the expected direction in most of the cities
in our sample, with a mean elasticity of around 3 that can structurally
be interpreted as households spending on average one-third of their
income net of transportation costs on housing. However, low data quality
or a high percentage of informal housing make the rent gradient flatter,
so that an important direction for future research should be to adapt
the monocentric SUM to such local contexts.

\subsection{Density gradients \label{subsec:Density-elasticities}}

\subsubsection{Empirical strategy}

This subsection examines a second relationship derived from the SUM:
the negative population density gradient from the city center to the
periphery. We expect population density to decrease with transportation
costs or increase with income net of transportation costs. Building
on equation \ref{eq:eq_density_distance-1-1-1}, we run the following
regression on the sampled cities:

\begin{equation}
ln(n_{i,c}){\displaystyle =\,}g_{c}+h_{c}ln(Y_{c}-T_{i,c})+\text{\textrtailn}_{i,c}\label{eq:empirical_density}
\end{equation}

We run this regression 192 times to individually estimate $g_{c}$
and $h_{c}$ for all the sampled cities. The error term $\lyxmathsym{\textrtailn}_{i,c}$
accounts for the fact that locations within cities differ in attributes
other than transportation costs. We can structurally interpret parameters
$h_{c}$ as $(1/\beta_{c}a_{c})-1$.

We empirically measure the income net of transportation costs as in
the previous subsection and use the Euclidean distance to the city
center as an instrumental variable. Population density in city \textit{c}
and pixel i is measured as $n_{i,c}=\frac{N_{i,c}}{L_{i,c}}$ with
$N_{i,c}$ the total population in the pixel derived from the GHSL
data and $L_{i,c}$ the amount of urbanizable land in the pixel derived
from ESA CCI land cover data.

\subsubsection{Results}

We start by analyzing the sign of parameter $h_{c}$. Based on the
theory, we expect this parameter to be positive: population density
should decrease when transportation costs increase, or when income
net of transportation costs decreases. Empirically estimating this
parameter in our sample, we find that it is significantly positive
at the 10\% level in all the cities.

Table \ref{tab:r2_density} shows the distribution of the R\textsuperscript{2}
of regression \ref{eq:empirical_density} in the sampled cities, which
can be interpreted as the percentage of the variance in population
densities explained by accessibility. We find a median R\textsuperscript{2}
of 0.17 and a maximum of 0.46, meaning that accessibility is a strong
determinant of population densities in a large number of cities in
our sample. These R\textsuperscript{2} figures are not as high as
in some existing studies (such as \citet{bertaud_spatial_2014}):
an explanation is that we analyze our data at a highly disaggregated
level, not grouping them by rings for instance \citep{duranton_chapter_2015}.\footnote{Running the same regressions, aggregating the data into 1km rings,
leads to a median R\textsuperscript{2} of 0.71 and a mean R\textsuperscript{2}
of 0.73 (OLS).}

\begin{table}[H]
\centering%
\begin{tabular}{cccccccc}
 & Mean & Min. & Q1 & Med. & Q3 & Max. & Nb. of obs.\tabularnewline
\hline 
OLS & 0.19 & 0.02 & 0.12 & 0.17 & 0.24 & 0.46 & 192\tabularnewline
2SLS & 0.15 & -1.12 & 0.09 & 0.15 & 0.23 & 0.46 & 192\tabularnewline
\hline 
\end{tabular}\caption{Distribution of the R\protect\textsuperscript{2} of equation \ref{eq:empirical_density}.\label{tab:r2_density}}
\end{table}

\begin{figure}[H]
\centering

\subfloat[Histogram - distribution of parameter \textit{h}\protect\textsubscript{\textit{c}}\textit{.}]{\centering\includegraphics[scale=0.7]{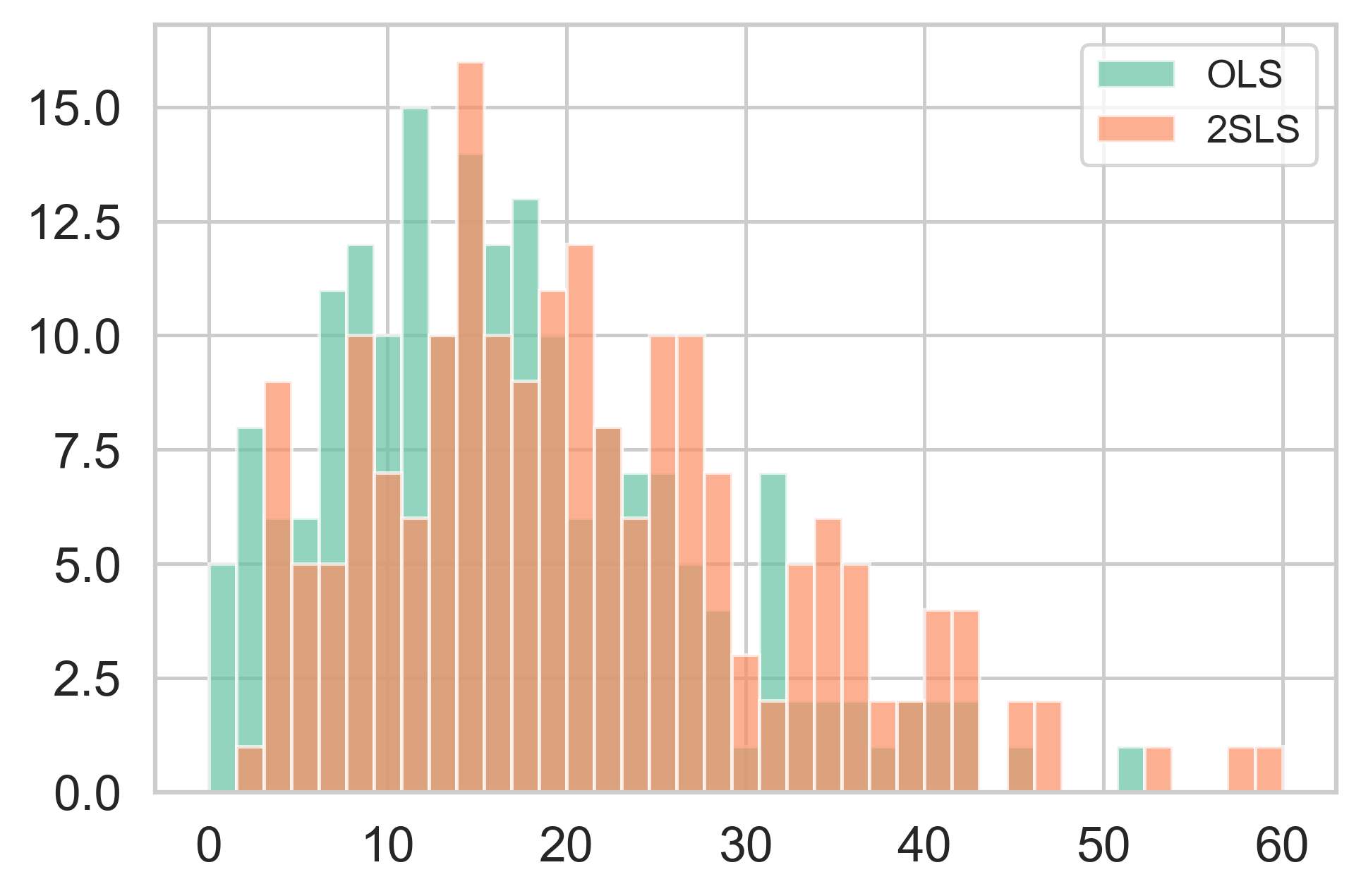}

}

\subfloat[Table - distribution of parameter \textit{h}\protect\textsubscript{\textit{c}}\textit{.}]{\centering%
\begin{tabular}{cccccccc}
 & Mean & Min. & Q1 & Med. & Q3 & Max. & Nb. of obs.\tabularnewline
\hline 
OLS & 16.70 & 0.84 & 9.28 & 15.19 & 22.33 & 50.94 & 192\tabularnewline
2SLS & 21.17 & 2.50 & 13.19 & 19.48 & 27.46 & 58.68 & 192\tabularnewline
\hline 
\end{tabular}

}\caption{Distribution of the estimates of parameter \textit{h}\protect\textsubscript{\textit{c}}
(elasticity of population density with respect to the income net of
transportation costs) in the 192 cities.\label{fig:distrib_dens}}
\end{figure}

Figure \ref{fig:distrib_dens} shows the distribution of the estimates
of parameter $h_{c}$ in the sampled cities. The mean estimate is
21.17, meaning that, on average, in our sample, when income net of
transportation costs decreases by 1\%, population density decreases
by 21.17\%. However, despite being always positive, estimates are
highly heterogeneous between cities. The minimum is 2.5, the maximum
58.7, and the interquartile range 13.2 - 27.5 (2SLS).

We can structurally derive the elasticity of housing production with
respect to land and capital (parameters \textit{$a_{c}$} and \textit{$b_{c}$}
from equation \ref{eq:housing_production_function}). From equation
\ref{eq:eq_density_distance-1-1-1}, \textit{$h_{c}=\frac{1-\beta_{c}a_{c}}{\beta_{c}a_{c}}$},
and from equation \ref{eq:rents}, \textit{$f_{c}=\frac{1}{\beta_{c}}$},
meaning that we can derive \textit{$a_{c}$} as $\frac{f_{c}}{1+h_{c}}$.
Using the estimates of $f_{c}$ from subsection \ref{subsec:Rent-elasticities}
and the estimates of $h_{c}$ from this subsection, we find that the
median estimate of the elasticity of housing production with respect
to land \textit{$a_{c}$} over the sample of cities is 0.13 (interquartile
range of 0.08-0.21) and that the median estimate of the elasticity
of housing production with respect to capital \textit{$b_{c}$} over
the sample of cities is 0.87 (interquartile range of 0.79 - 0.92).
Our results are in line with \citet{epple_new_2010}, who estimate
a land elasticity of 0.14, but slightly differ from \citet{combes_production_2021},
who estimate a capital elasticity of 0.65 for France.

\subsubsection{Second-step regressions}

This section further investigates the heterogeneity in density gradients
by regressing our estimates of \textit{h}\textsubscript{\textit{c}}
against city characteristics (Table \ref{tab:second_step_density}).

Specification (1) is based on the density gradient literature, which
suggests that polycentrism leads to flatter density gradients. Some
studies show that larger populations lead to flatter gradients, an
explanation being that employment subcenters are more likely to emerge
in larger urban areas (see subsection \ref{subsec:ev-Rent-gradients}).
Incomes and transportation costs have also been found to play a role
in explaining density gradients (see subsection \ref{subsec:ev-Rent-gradients}).
However, they are already accounted for in our first step empirical
regressions (equation \ref{eq:empirical_density}), and therefore
we do not need to include them in this second-step regression.

Empirically testing these hypotheses, we find that larger (in terms
of population) and more polycentric urban areas tend to have flatter
density gradients. The R\textsuperscript{2} of this first specification
is 0.38, meaning that population and polycentricity explain a large
percentage of the variance in density gradients across urban areas.
Since, here, population is included in the second-step regression
as an indicator of polycentricity, it may be surprising that population
and monocentricity are both significant. We attribute this to the
quality of our monocentricity index: it indicates whether the historical,
geographical, transportation, administrative, and business centers
are all located in the same place, but it does not indicate whether
the city has a large number of subcenters, which is proxied by the
city's population. Thus, the total population and the monocentricity
index correspond to two different definitions of monocentricity.

In specification (2), we add four explanatory variables: the level
of inequalities, the percentage of informal housing, and two categorical
variables indicating whether the city is coastal and whether there
is a stringent regulatory environment. The first two variables are
based on our observation that rent gradients seem to be flattened
in cities with a large percentage of informal housing or inequalities
(Figure \ref{fig:sign_rent_grad}). The two categorical variables
are based on \citet{bertaud_spatial_2014}, who find that complex
geographies or stringent regulatory environments flatten the density
gradient.We find that a large percentage of informal housing and high
inequality levels tend to flatten the population density gradient,
the two variables being significant at the 1\% level. Monocentricity
and population remain significant as well.

Specification (3) adds continent fixed effects, taking Europe as the
reference. Population, monocentricity, and informal housing remain
significantly different from 0 at the 1\%, 1\%, and 5\% levels respectively:
cities with larger populations, more subcenters, and more informal
housing tend to have flatter density gradients. Surprisingly, the
dummy for stringent regulatory environments is positive and significant
at the 10\% level, meaning that cities with strong regulations tend
to have steeper density gradients, which contrasts with \citet{bertaud_spatial_2014}.
However, as \citet{bertaud_spatial_2004} has observed, some cities
subject to stringent regulations developed before the implementation
of the latter, and have kept an urban form that is to some extent
similar to market-oriented cities. For instance, Central and Eastern
European cities, such as Cracow, were stringently regulated during
the socialist era, but these cities had a large historical center
established many centuries before socialism so that they have remained
largely monocentric. Finally, African and Oceanian cities tend to
have flatter density gradients.

\begin{table}[H]
\centering%
\begin{tabular}{cccc}
\hline 
 & \multicolumn{3}{c}{\textit{Dependent variable: population density gradients h}\textsubscript{\textit{c}}}\tabularnewline
\cline{2-4} \cline{3-4} \cline{4-4} 
 &  &  & \tabularnewline
 & (1) & (2) & (3)\tabularnewline
\hline 
 &  &  & \tabularnewline
log(population) & -5.895$^{***}$ & -4.192$^{***}$ & -5.428$^{***}$\tabularnewline
 & (0.626) & (0.690) & (0.856)\tabularnewline
Monocentricity index & 6.525$^{***}$ & 3.148$^{**}$ & 4.346$^{***}$\tabularnewline
 & (1.314) & (1.463) & (1.591)\tabularnewline
Gini index &  & -0.263$^{***}$ & -0.104\tabularnewline
 &  & (0.087) & (0.147)\tabularnewline
Percentage of informal housing &  & -0.218$^{***}$ & -0.176$^{**}$\tabularnewline
 &  & (0.053) & (0.070)\tabularnewline
Coastal city &  & 0.386 & 1.583\tabularnewline
 &  & (1.362) & (1.508)\tabularnewline
Regulatory environment &  & 2.512 & 3.119$^{*}$\tabularnewline
 &  & (1.584) & (1.749)\tabularnewline
Asia &  &  & 2.602\tabularnewline
 &  &  & (3.939)\tabularnewline
Africa &  &  & -8.157$^{***}$\tabularnewline
 &  &  & (2.968)\tabularnewline
Oceania &  &  & -9.958$^{***}$\tabularnewline
 &  &  & (2.442)\tabularnewline
North America &  &  & 2.598\tabularnewline
 &  &  & (3.429)\tabularnewline
South America &  &  & -3.264\tabularnewline
 &  &  & (3.801)\tabularnewline
Constant & 92.486$^{***}$ & 86.716$^{***}$ & 95.918$^{***}$\tabularnewline
 & (10.650) & (11.113) & (12.958)\tabularnewline
\hline &  &  & \tabularnewline
Observations & 192 & 192 & 192\tabularnewline
$R^{2}$ & 0.375 & 0.448 & 0.508\tabularnewline
Adjusted $R^{2}$ & 0.368 & 0.430 & 0.478\tabularnewline
F Statistic & 70.21$^{***}$ & 40.15$^{***}$ & 29.49$^{***}$\tabularnewline
\hline\hline &  &  & \tabularnewline
\textit{Note:} & \multicolumn{3}{c}{$^{*}$p$<$0.1; $^{**}$p$<$0.05; $^{***}$p$<$0.01}\tabularnewline
\end{tabular}

\caption{Second-step regression of population density gradient estimates \textit{h}\protect\textsubscript{\textit{c}}
against city characteristics.\label{tab:second_step_density}}
\end{table}

\subsubsection{Robustness checks}

We run some robustness checks for alternative specifications of equation
\ref{eq:empirical_density}. In particular, we regress the logarithm
of population density on the Euclidean distance from the city center,
instead of considering the logarithm of the income net of transportation
costs. On average, in the sample of cities, we find a gradient of
-0.085, meaning that going 1km farther from the city center decreases
population density by 8.5\% (\ref{sec:Robustness-checks}, Figure
\ref{fig:distrib_density_robustness}). We find some heterogeneity
in the estimates of this gradient between cities: the interquartile
range is -0.102 - -0.060, and results of the second-step regression
show that this gradient is flatter in North America, Oceania, and
when the city is larger in terms of population, and steeper when the
city is coastal or has a large percentage of informal housing.

We run some additional robustness checks, including regressing the
logarithm of population densities on the logarithm of transportation
costs or transportation times instead of regressing them on the logarithm
of incomes nets of transportation costs. Full details and results
are in \ref{sec:Robustness-checks}. Overall, our results are robust
to alternative specifications. A density gradient of the expected
direction is found in all of the cities and with all robustness checks.
The population of the city, its monocentricity, and continent-fixed
effects are found to impact the value of the density gradient in most
of the robustness checks. The percentage of informal housing has an
impact on density gradients in some of the robustness checks performed.

\medskip{}
To summarize this subsection, we find that the expected negative population
density gradient is observed in all of the cities in our sample. In
addition, in line with the literature, we found flatter density gradients
in larger, and more polycentric, cities. A direction for future research
should be incorporating informality into urban models, as we also
observed flatter density gradients in cities with large percentages
of informal housing.

\subsection{Urban area}

This subsection examines whether the size of urban areas increases
with population and income and decreases with farmland rents and transportation
costs, as suggested by the SUM (equation \ref{eq:urban_area_eq}).

In the first specification (``all cities'') of Table \ref{tab:urban_area},
we regress urbanized areas on a range of city characteristics derived
from the theory (subsection \ref{subsec:Theory}), including populations,
incomes, and farmland rents. We add two commuting cost variables:
fuel price and \textquotedbl commuting speed\textquotedbl , which
is the density-weighted average speed when traveling to the city center
during the evening rush hour using the fastest transportation mode
(car or public transport). We also include the monocentricity index,
as many studies argue that polycentricity impacts the size of urban
areas. \citet{spivey_millsmuth_2008} finds that more sub-centers
are associated with a smaller land area in the United States, which
she interprets as the fact that sub-centers are more likely to develop
in dense areas or that increased density associated with sub-centers
mitigates sprawl. \citet{paulsen_yet_2012} confirms this result.
In contrast, \citet{burchfield_causes_2006} find that decentralized
employment increases sprawl in the US. \citet{schmidt_does_2020}
find non-significant estimates of the impact of polycentricity on
the size of urban areas in Germany.

We find an R\textsuperscript{2} of 0.86, meaning that city characteristics
explain a large percentage of the variance in the size of urbanized
areas among the sampled cities. In addition, all the coefficients
of the explanatory variables are significant and of the expected sign,
except for fuel price. For instance, a 1\% increase in population
leads to a 0.86\% increase in the urban area, in line with \citet{spivey_millsmuth_2008},
which finds an elasticity of 0.91, and slightly higher than \citet{paulsen_yet_2012},
which finds elasticities of between 0.56 and 0.64. We also find that
a 1\% increase in incomes leads to a 0.40\% increase in the urban
area. An increase in farmland rents tends to decrease urban areas,
and an increase in commuting speed, corresponding to a decrease in
the opportunity cost of transportation, increases urban areas. The
non-significant estimate for fuel price might be caused by cities'
inertia when compared to the relative volatility of fuel prices. Finally,
for monocentricity, we find a positive coefficient significant at
the 10\% level, in line with \citet{spivey_millsmuth_2008}.

\begin{table}[H]
\centering%
\begin{tabular}{@{\extracolsep{5pt}}lccc}
 &  &  & \tabularnewline
 &  &  & \tabularnewline
\hline 
 & \multicolumn{3}{c}{\textit{Dependent variable: log(urbanized area)}}\tabularnewline
\cline{2-4} &  &  & \tabularnewline
 & \multirow{2}{*}{All cities} & \multirow{2}{*}{High income} & Upper-middle income\tabularnewline
 &  &  & to low income\tabularnewline
\hline &  &  & \tabularnewline
log(population) & 0.856$^{***}$ & 0.913$^{***}$ & 0.778$^{***}$\tabularnewline
 & (0.040) & (0.067) & (0.047)\tabularnewline
log(income) & 0.397$^{***}$ & 0.277$^{**}$ & 0.436$^{***}$\tabularnewline
 & (0.038) & (0.125) & (0.055)\tabularnewline
log(farmland rents) & -0.220$^{***}$ & -0.311$^{***}$ & -0.076\tabularnewline
 & (0.043) & (0.056) & (0.047)\tabularnewline
log(fuel price) & -0.061 & 0.125 & -0.153\tabularnewline
 & (0.101) & (0.213) & (0.136)\tabularnewline
log(commuting speed) & 0.466$^{***}$ & 0.434$^{**}$ & 0.563$^{***}$\tabularnewline
 & (0.100) & (0.168) & (0.132)\tabularnewline
Monocentricity index & 0.146$^{*}$ & 0.101 & 0.199$^{**}$\tabularnewline
 & (0.082) & (0.172) & (0.085)\tabularnewline
Constant & -12.876$^{***}$ & -12.413$^{***}$ & -12.070$^{***}$\tabularnewline
 & (0.624) & (0.998) & (0.847)\tabularnewline
\hline &  &  & \tabularnewline
Observations & 192 & 108 & 84\tabularnewline
$R^{2}$ & 0.858 & 0.866 & 0.878\tabularnewline
Adjusted $R^{2}$ & 0.853 & 0.858 & 0.869\tabularnewline
F Statistic & 225.7$^{***}$ & 184.9$^{***}$ & 96.78$^{***}$\tabularnewline
\hline\hline &  &  & \tabularnewline
\textit{Note:} & \multicolumn{3}{r}{$^{*}$p$<$0.1; $^{**}$p$<$0.05; $^{***}$p$<$0.01}\tabularnewline
\end{tabular}\caption{Regression of urbanized areas against city characteristics.\label{tab:urban_area}}
\label{tab:aggregated}
\end{table}

To further investigate the heterogeneity in the determinants of urbanized
areas between cities, we slice the result by income levels. “High
income” cities are those located in high-income countries, according
to the World Bank classification. “Upper-middle income to low income”
corresponds to the cities located in upper-middle, lower-middle, or
low income countries according to the World Bank classification. In
our sample, all North American and Oceanian cities are classified
as high-income, as well as 2 Asian cities (Hong Kong and Singapore),
4 South American cities (in Uruguay and Chile), and most European
cities (83/97). The remaining cities are all African cities, most
Asian cities (30/32), and most South American cities (30/34). This
classification allows us to split our sample into two groups each
containing a similar number of cities, with each group broadly corresponding
to three continents. Unfortunately, we cannot slice the results by
continent due to data limitations: we do not have enough cities for
some continents (we only have 8 cities for Oceania and 10 cities for
Africa) or enough heterogeneity within continents (the fact that all
of the North American cities of the sample are located in the USA
is problematic because we only have data at the country level for
some explanatory variables).

We run a Chow test to check whether we need separate models for these
two groups of cities. The test is significant at the 5\% level, meaning
that the differences in estimated coefficients across the two sub-samples
are not the result of sampling error.

We find that an increase in population has a larger impact on the
size of urbanized areas in high-income cities than in other cities,
whereas income has a larger impact on the size of urbanized areas
in middle and low-income cities. The latter result is in line with
the studies that find that income growth is a large driver of urban
sprawl in developing countries \citep{deng_growth_2008,ke_determinants_2009}.
It is also in line with \citet{schmidt_does_2020}: they find that
income has a low impact on urban sprawl in Germany, which they explain
by the growth management policies that eliminate the impact of income
in Germany. Farmland rents have a large negative impact on urban sprawl
in high-income countries, in line with the findings of \citet{oueslati_determinants_2015}:
they found a high coefficient for farmland rents, which, according
to the authors, reflects the fact that in Europe, agriculture on the
urban fringe is often highly intensive, offering relatively high yields
and profits. Commuting speed has a smaller impact on high-income countries.
Finally, monocentricity is non-significant in high-income countries,
whereas it is significant for middle-income and low-income countries.

\section{Conclusion\label{section:discussion}}

Exploiting a new spatially-explicit dataset on population densities,
land cover, but also real estate and transportation costs, we are
able to empirically assess the main relationships derived from the
SUM in a large and diverse sample of cities. In particular, we can
assess the SUM's predictions in terms of internal city structures,
evaluating density and rent gradients in 192 global cities.

Although the SUM is an old, simple model, we show that its main predictions
are verified for our sample of cities. We find the expected declining
rent gradient from city centers to suburbs in 167 out of 192 cities
and the expected declining density gradient from city centers to suburbs
in all the sampled cities. In addition, we find that the total urbanized
area of a city tends to increase with population and income and decrease
with farmland rents and transportation costs, as expected from the
theory.

Our dataset also allows us to investigate heterogeneity in the internal
structures of cities. Large or polycentric cities are more likely
to have flat density gradients, as are cities with a higher percentage
of informal housing. Cities are more likely to have flat or inverted
rent gradients if they have a large percentage of informal housing
or if the quality of our rent data is lower. Overall, urban sprawl
in high-income cities is largely driven by population and farmland
rents, whereas income plays a key role in developing countries.

This result implies that the SUM does not perfectly fit all cities.
Coastal amenities, polycentricity, and informal housing can lead to
flatter, and sometimes inverted, rents and density gradients, so that
modelling cities more accurately requires more complex models that
account for these dimensions, in particular beyond the SUM's traditional
application area (North America, Europe).

A first limitation of this study is that it only assesses the simplest
version of the SUM. Indeed, data were missing that would have enabled
systematic study of more advanced urban models. Gathering data on
employment subcenters and transportation costs to these employment
centers is, for instance, outside the scope of this study. However,
the development of new big data approaches will allow this gap to
be filled in future research \citep{barzin_where_2022}.

Another limitation is that the SUM is mostly a static model, whereas
there is inertia in cities' development. Densities and built environment
are changing slowly (as are transport times, to some extent), whereas
rents and land prices can change quickly. Obtaining panel data on
cities would be necessary in order to study changes in city structures
over time.

\section*{Acknowledgments}

This study was financed by the Dragon Project (ANR-14-ORAR-0005).
We thank Amandine Toussaint and Basile Pfeiffer for interesting insights
and discussions during the framing of this study, and Felix Creutzig
as well as participants at the 15th North American Meeting of the
Urban Economics Association for their comments and advice. We also
gratefully thank the editor Gabriel Ahlfeldt and two anonymous reviewers
for their insightful and critical comments, which have significantly
improved the quality of the manuscript.

\section*{References}

\bibliographystyle{chicago}
\bibliography{SUM_validation}

\newpage{}

\appendix

\section{Land cover classification\label{appendix:land_cover}}

\begin{table}[H]
\small\scalebox{0.8}{%
\begin{tabular}{ll}
\toprule 
ESA CCI land cover category & Reclassification\tabularnewline
\midrule 
10 Cropland, rainfed & Urbanizable\tabularnewline
11 Herbaceous cover & Urbanizable\tabularnewline
12 Tree or shrub cover & Urbanizable\tabularnewline
20 Cropland, irrigated or post-flooding & Urbanizable\tabularnewline
30 Mosaic cropland (>50\%) / natural vegetation (<50\%) & Urbanizable\tabularnewline
40 Mosaic natural vegetation (>50\%) / cropland (<50\%) & Urbanizable\tabularnewline
50 Tree cover, broadleaved, evergreen, closed to open (>15\%) & Urbanizable\tabularnewline
60 Tree cover, broadleaved, deciduous, closed to open (>15\%) & Urbanizable\tabularnewline
61 Tree cover, broadleaved, deciduous, closed (>40\%) & Urbanizable\tabularnewline
62 Tree cover, broadleaved, deciduous, open (15-40\%) & Urbanizable\tabularnewline
70 Tree cover, needleleaved, evergreen, closed to open (>15\%) & Urbanizable\tabularnewline
71 Tree cover, needleleaved, evergreen, closed (>40\%) & Urbanizable\tabularnewline
72 Tree cover, needleleaved, evergreen, open (15-40\%) & Urbanizable\tabularnewline
80 Tree cover, needleleaved, deciduous, closed to open (>15\%) & Urbanizable\tabularnewline
81 Tree cover, needleleaved, deciduous, closed (>40\%) & Urbanizable\tabularnewline
82 Tree cover, needleleaved, deciduous, open (15-40\%) & Urbanizable\tabularnewline
90 Tree cover, mixed leaf type (broadleaved and needleleaved) & Urbanizable\tabularnewline
100 Mosaic tree and shrub (>50\%) / herbaceous cover (<50\%) & Urbanizable\tabularnewline
110 Mosaic herbaceous cover (>50\%) / tree and shrub (<50\%) & Urbanizable\tabularnewline
120 Shrubland & Urbanizable\tabularnewline
121 Evergreen shrubland & Urbanizable\tabularnewline
122 Deciduous shrubland & Urbanizable\tabularnewline
130 Grassland & Urbanizable\tabularnewline
140 Lichens and mosses & Urbanizable\tabularnewline
150 Sparse vegetation (tree, shrub, herbaceous cover) (<15\%) & Urbanizable\tabularnewline
160 Tree cover, flooded, fresh or brackish water & Non urbanizable\tabularnewline
170 Tree cover, flooded, saline water & Non urbanizable\tabularnewline
180 Shrub or herbaceous cover, flooded, fresh/saline/brakish water & Non urbanizable\tabularnewline
190 Urban areas & Urbanizable\tabularnewline
200 Bare areas & Urbanizable\tabularnewline
201 Consolidated bare areas & Urbanizable\tabularnewline
202 Unconsolidated bare areas & Urbanizable\tabularnewline
210 Water bodies & Non urbanizable\tabularnewline
220 Permanent snow and ice & Non urbanizable\tabularnewline
\bottomrule
\end{tabular}}\caption{Reclassification of the ESA CCI land cover categories into \textquotedbl urbanizable\textquotedbl{}
and \textquotedbl non urbanizable\textquotedbl}
\end{table}

\section{Robustness checks\label{sec:Robustness-checks}}

\subsection{Description of the robustness checks}

We run five robustness checks, detailed below, for subsections \ref{subsec:Rent-elasticities}
and \ref{subsec:Density-elasticities}. In robustness check 1, our
variable of interest is the Euclidean distance to the city center
$d_{i,c}$ instead of the income net of transportation costs $Y_{c}-T_{i,c}$.
Our theoretical model is similar to subsection \ref{subsec:Theory},
except that households seek to maximize the following utilities:

\begin{equation}
max\hspace{3mm}U(z_{i,c},\,q_{i,c},d_{i,c})=z_{i,c}^{\alpha_{c}}\,q_{i,c}^{\beta_{c}}l(d_{i,c})\quad s.t.\hspace{3mm}z_{i,c}+q_{i,c}R_{i,c}\leq Y_{c}\label{eq:max-1-1}
\end{equation}
with $l$ an amenity that depends on the Euclidean distance to the
city center $d_{i,c}$. From equation \ref{eq:max-1-1}, we find that:
\begin{equation}
\begin{array}{l}
q_{i,c}=\beta_{c}Y_{c}/R_{i,c}\\
z_{i,c}=\alpha_{c}Y_{c}
\end{array}\label{eq:dwelling_size-1}
\end{equation}
Using the spatial equilibrium condition and writing $u_{c}$ the utility
at equilibrium: 
\begin{equation}
R_{i,c}=R_{0,c}\,l(d_{i,c})^{1/\beta_{C}}\hspace{1cm}\text{with}\hspace{1cm}R_{0,c}=(\frac{\alpha_{c}^{\alpha_{c}}\,\beta_{c}^{\beta_{c}}\,Y_{c}}{u_{c}})^{1/\beta_{c}}\label{eq:rents-1}
\end{equation}
The housing production function remains unchanged from section \ref{subsec:Theory}:
\begin{equation}
H(K_{i,c},\,L_{i,c})=A_{c}\,L_{i,c}^{a_{c}}\,K_{i,c}^{b_{c}}\label{eq:housing_production_function-1}
\end{equation}
As in subsection \ref{subsec:Theory}, from the profit maximization
of private developers, we can derive the total population per pixel
$N_{i,c}$:
\begin{equation}
N_{i,c}=H_{i,c}/q_{i,c}=A_{c}^{1/a_{c}}(b_{c}R_{i,c}/\rho)^{b_{c}/a_{c}}L_{i,c}/q_{i,c}\label{eq:-1-1}
\end{equation}
Replacing $q_{i,c}$ and $R_{i,c}$ by their expressions in equations
\ref{eq:dwelling_size-1} and \ref{eq:rents-1}, the population density
par pixel $n_{i,c}=N_{i,c}/L_{i,c}$ can be expressed as:

\begin{equation}
n_{i,c}=A_{c}^{1/a_{c}}(\frac{b_{c}}{\rho})^{b_{c}/a_{c}}(\frac{\alpha_{c}^{\alpha_{c}}}{u})^{1/(\beta_{c}a_{c})}\beta_{c}^{b_{c}/a_{c}}Y_{c}^{(1-a_{c}\beta_{c})/a_{c}\beta_{c}}l(d_{i,c})^{1/a_{c}\beta_{c}}\label{eq:eq_density_distance-1-1-1-1}
\end{equation}
Our empirical strategy is then as follows: assuming that $l(d_{i,c})=\kappa e^{-\theta d_{i,c}}$and
building on equations \ref{eq:rents-1} and \ref{eq:eq_density_distance-1-1-1-1},
we run the following regressions on the cities in the database:

\begin{equation}
ln(R_{i,c})=e'_{c}+f'_{c}d_{i,c}+\varepsilon'_{i,c}
\end{equation}

\begin{equation}
ln(n_{i,c})=g'_{c}+h'_{c}d_{i,c}+\eta'_{i,c}
\end{equation}
$e'_{c}$ and $g'_{c}$ are constants, $\varepsilon'_{c}$ and $\eta'_{c}$
are error terms, and $f'_{c}$ and $h'_{c}$ can be structurally interpreted
as $-\theta/\beta_{c}$ and $\frac{-\theta}{a_{c}\beta_{c}}$ respectively.

\medskip{}

In addition to robustness check 1, we run four additional robustness
checks:
\begin{itemize}
\item Robustness check 2: We regress the logarithm of population densities
and rents on the logarithm of transportation costs, instead of the
logarithm of income net of transportation costs. We substitute $T_{i,c}$
for $Y_{c}-T_{i,c}$ in equations \ref{eq:empirical_rent} and \ref{eq:empirical_density}.
\item Robustness check 3: We regress the logarithm of population densities
and rents on the logarithm of transportation times (taking the minimum
of the transportation time by driving and public transport). We substitute
the minimum of the transportation time by driving and public transport
$t_{i,c}=min(t_{i,c}^{driving},t_{i,c}^{publictransport})$ for $Y_{c}-T_{i,c}$
in equations \ref{eq:empirical_rent} and \ref{eq:empirical_density}.
\item Robustness check 4: We regress the logarithm of population densities
and rents on the logarithm of incomes net of transportation costs,
but omitting the monetary cost of public transport as data are not
consistently available for all cities.
\item Robustness check 5: We regress the logarithm of population densities
and rents on the logarithm of incomes net of transportation costs
(equation \ref{eq:empirical_rent}). However, instead of taking the
average rent per grid cell, rent data are aggregated by regressing
rents on dwelling sizes in each grid cell (this robustness check is
relevant for results regarding rent gradients, but does not affect
the results regarding density gradients).
\end{itemize}

\subsection{Robustness checks - Rent gradients}

With our main specification (subsection \ref{subsec:Rent-elasticities}),
we find rent gradients of the expected direction in 160 cities with
OLS and 167 with 2SLS (Figure \ref{fig:sign_rent_grad}). We investigate
whether this result holds under the five robustness checks (Table
\ref{tab:robustness_rent_directop,}).

\begin{table}[H]
\begin{centering}
\begin{tabular}{|c|c|c|c|c|}
\hline 
 &  & Expected & Opposite & Non-significant\tabularnewline
 &  & direction & direction & (10\% level)\tabularnewline
\hline 
Main specification & OLS & 160 & 29 & 2\tabularnewline
Main specification & 2SLS & 167 & 23 & 1\tabularnewline
Robustness check 1 & OLS & 160 & 29 & 2\tabularnewline
Robustness check 2 & OLS & 163 & 27 & 1\tabularnewline
Robustness check 2 & 2SLS & 167 & 23 & 1\tabularnewline
Robustness check 3 & OLS & 162 & 28 & 1\tabularnewline
Robustness check 3 & 2SLS & 167 & 23 & 1\tabularnewline
Robustness check 4 & OLS & 159 & 30 & 2\tabularnewline
Robustness check 4 & 2SLS & 167 & 23 & 1\tabularnewline
Robustness check 5 & OLS & 157 & 32 & 2\tabularnewline
Robustness check 5 & 2SLS & 161 & 29 & 1\tabularnewline
\hline 
\end{tabular}\caption{Robustness checks -- direction of rent gradients.\label{tab:robustness_rent_directop,}}
\par\end{centering}
\end{table}

With our main specification (subsection \ref{subsec:Rent-elasticities}),
we find that the median R\textsuperscript{2} of regression \ref{eq:empirical_rent}
is 0.18 with OLS and 0.16 with 2SLS (Table \ref{tab:r2_rent}). We
investigate whether this result holds under the five robustness checks
(Table \ref{tab:robustness_rent_r2}).

\begin{table}[H]
\begin{centering}
\begin{tabular}{|c|c|c|c|c|c|c|c|}
\hline 
 &  & Mean & Min. & Q1 & Med. & Q3 & Max.\tabularnewline
\hline 
Main specification & OLS & 0.20 & 0.00 & 0.06 & 0.18 & 0.31 & 0.66\tabularnewline
Main specification & 2SLS & 0.19 & -0.17 & 0.04 & 0.16 & 0.30 & 0.66\tabularnewline
Robustness check 1 & OLS & 0.20 & 0.00 & 0.06 & 0.17 & 0.31 & 0.61\tabularnewline
Robustness check 2 & OLS & 0.20 & 0.00 & 0.07 & 0.18 & 0.32 & 0.64\tabularnewline
Robustness check 2 & 2SLS & 0.20 & -0.00 & 0.07 & 0.17 & 0.31 & 0.64\tabularnewline
Robustness check 3 & OLS & 0.19 & 0.00 & 0.06 & 0.17 & 0.29 & 0.63\tabularnewline
Robustness check 3 & 2SLS & 0.19 & -0.02 & 0.06 & 0.16 & 0.27 & 0.62\tabularnewline
Robustness check 4 & OLS & 0.19 & 0.00 & 0.06 & 0.17 & 0.30 & 0.59\tabularnewline
Robustness check 4 & 2SLS & 0.17 & -0.17 & 0.04 & 0.16 & 0.29 & 0.56\tabularnewline
Robustness check 5 & OLS & 0.18 & 0.00 & 0.04 & 0.15 & 0.28 & 0.66\tabularnewline
Robustness check 5 & 2SLS & 0.17 & -0.13 & 0.03 & 0.14 & 0.27 & 0.66\tabularnewline
\hline 
\end{tabular}\caption{Robustness checks - R\protect\textsuperscript{2} of equation \ref{eq:empirical_rent}.\label{tab:robustness_rent_r2}}
\par\end{centering}
\end{table}

Then, we display the distribution of the rent gradients estimated
in Robustness check 1. On average, moving 1km farther from the city
center reduces rents per m\textsuperscript{2} by 1.4\%.

\begin{figure}[H]
\centering

\subfloat[Histogram - distribution of parameter \textit{f'}\protect\textsubscript{\textit{c}}\textit{.}]{\centering\includegraphics[scale=0.7]{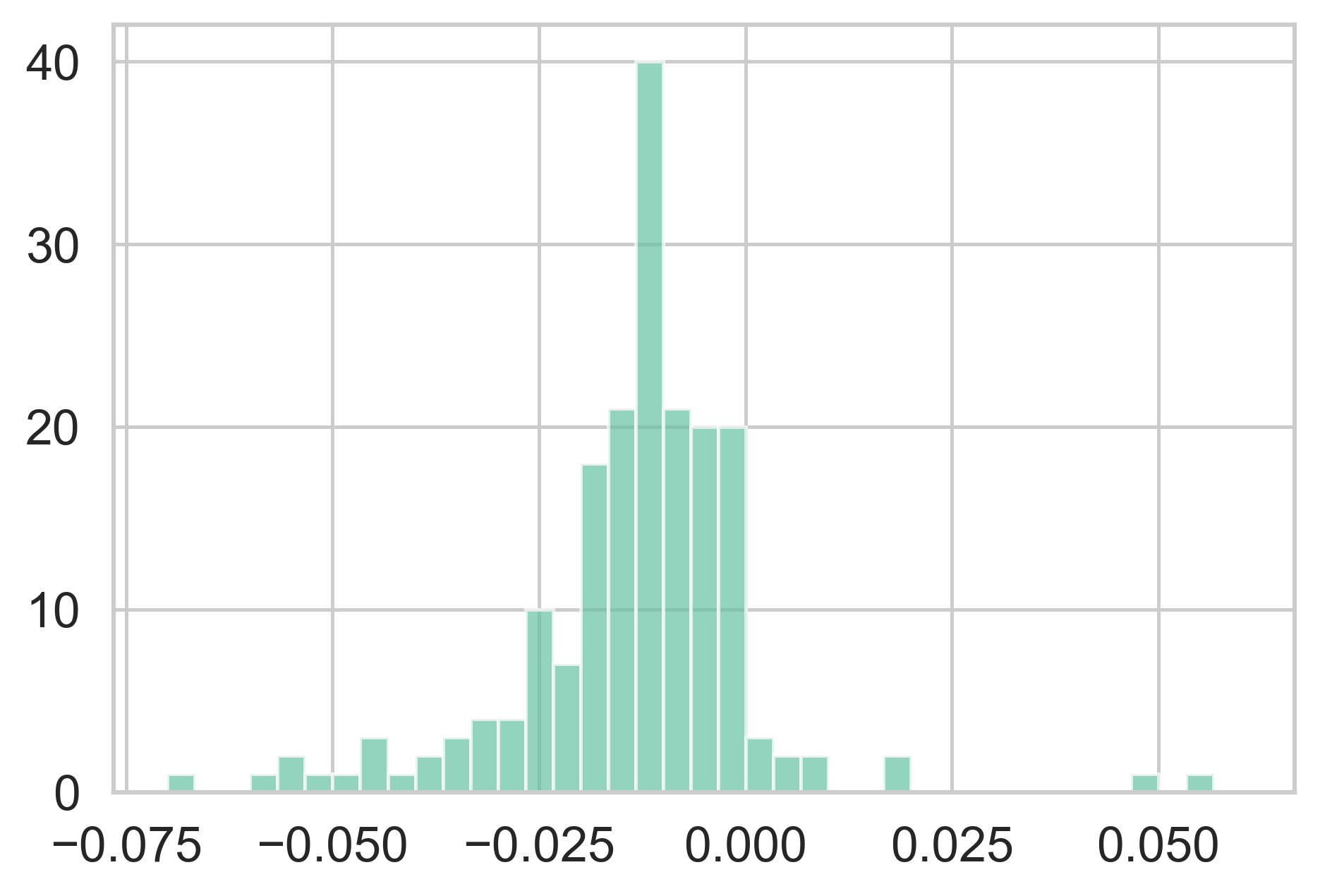}

}

\subfloat[Table - distribution of parameter \textit{f'}\protect\textsubscript{\textit{c}}\textit{.}]{\centering%
\begin{tabular}{cccccccc}
 & Mean & Min. & Q1 & Med. & Q3 & Max. & Nb. of obs.\tabularnewline
\hline 
OLS & -0.014 & -0.070 & -0.019 & -0.012 & -0.006 & 0.056 & 191\tabularnewline
\hline 
\end{tabular}

}\caption{Distribution of the estimates of parameter \textit{f'}\protect\textsubscript{\textit{c}}
(rent gradient estimated with Robustness check 1) in the 191 cities.\label{fig:distrib_rent_robustness}}
\end{figure}

With our main specification (subsection \ref{subsec:Rent-elasticities}),
we find that the market data cover, the percentage of informal housing,
and the dummy indicating whether the city is located in Asia explain
the estimated rent gradients (Table \ref{tab:rent_grad_second_step}).
We investigate whether this result holds under the five robustness
checks (Table \ref{tab:robustness_rent_second_step}).

\begin{table}[H]
\begin{centering}
\scalebox{0.9}{%
\begin{tabular}{ccccccc}
\hline 
 & \multicolumn{6}{c}{\textit{Dependent variable: rents gradients}}\tabularnewline
\cline{2-7} \cline{3-7} \cline{4-7} \cline{5-7} \cline{6-7} \cline{7-7} 
 & Main & Robustness & Robustness & Robustness & Robustness & Robustness\tabularnewline
 & specification & check 1 & check 2 & check 3 & check 4 & check 5\tabularnewline
\hline 
Coastal city & -0.395 & 0.003 & 0.016 & 0.001 & -0.352 & -0.530\tabularnewline
 & (0.409) & (0.002) & (0.028) & (0.033) & (0.414) & (0.432)\tabularnewline
Monocentricity index & 0.567 & -0.002 & 0.004 & 0.007 & 0.580 & 0.663\tabularnewline
 & (0.487) & (0.003) & (0.035) & (0.040) & (0.493) & (0.514)\tabularnewline
log(population) & -0.255 & 0.001 & -0.091{*}{*}{*} & -0.092{*}{*}{*} & -0.280 & -0.207\tabularnewline
 & (0.236) & (0.001) & (0.017) & (0.019) & (0.239) & (0.250)\tabularnewline
log(income) &  & 0.001 & -0.042 & -0.049 &  & \tabularnewline
 &  & (0.002) & (0.031) & (0.036) &  & \tabularnewline
Market data cover & -0.000{*}{*}  & 0.000{*}{*}{*} & 0.000{*}{*}{*} & 0.000{*}{*}{*} & -0.000{*}{*} & -0.000{*}{*}\tabularnewline
 & (0.000) & (0.000) & (0.000) & (0.000) & (0.000) & (0.000)\tabularnewline
Spatial data cover & 3.196 & 0.010 & 0.128 & 0.134 & 3.545 & 3.273\tabularnewline
 & (2.593) & (0.015) & (0.204) & (0.234) & (2.622) & (2.737)\tabularnewline
Gini index & -0.037 & 0.000 & 0.004 & 0.004 & -0.038 & -0.013\tabularnewline
 & (0.044) & (0.000) & (0.003) & (0.004) & (0.045) & (0.047)\tabularnewline
Informal housing & -0.091{*}{*}{*}  & 0.000{*}{*}{*} & 0.006{*}{*}{*} & 0.007{*}{*}{*} & -0.092{*}{*}{*} & -0.084{*}{*}{*}\tabularnewline
 & (0.026) & (0.000) & (0.002) & (0.002) & (0.026) & (0.027)\tabularnewline
Regulatory environment & 0.374 & -0.003 & -0.034 & -0.044 & 0.394 & -0.064\tabularnewline
 & (0.404) & (0.002) & (0.029) & (0.034) & (0.408) & (0.426)\tabularnewline
Asia & 4.065{*}{*}{*}  & -0.022{*}{*}{*} & -0.173{*}{*} & -0.024{*}{*}{*} & 4.041{*}{*}{*} & 3.999{*}{*}{*}\tabularnewline
 & (0.919) & (0.005) & (0.069) & (0.079) & (0.930) & (0.970)\tabularnewline
Africa & 0.329 & -0.002 & 0.025 & -0.012 & 0.203 & 0.454\tabularnewline
 & (1.213) & (0.007) & (0.093) & (0.107) & (1.227) & (1.280)\tabularnewline
Oceania & -1.275 & 0.004 & 0.015 & 0.040 & -1.309 & -0.950\tabularnewline
 & (0.994) & (0.005) & (0.070) & (0.080) & (1.006) & (1.049)\tabularnewline
North America & -0.060 & 0.002 & 0.131{*} & 0.140{*} & -0.210 & -0.096\tabularnewline
 & (0.946) & (0.005) & (0.071) & (0.078) & (0.998) & (0.999)\tabularnewline
South America & 1.429 & -0.014{*}{*}{*} & -0.091 & -0.117 & 1.346 & 1.135\tabularnewline
 & (0.987) & (0.005) & (0.071) & (0.081) & (0.998) & (1.042)\tabularnewline
Constant & 6.576{*}{*}  & -0.052{*} & 1.292{*}{*}{*} & 1.352{*}{*}{*} & 7.030{*}{*} & 4.482\tabularnewline
 & (3.213) & (0.030) & (0.407) & (0.467) & (3.249) & (3.391)\tabularnewline
\hline &  &  &  &  &  & \tabularnewline
Observations & 191 & 191 & 191 & 191 & 191 & 191\tabularnewline
R\textsuperscript{2}  & 0.227 & 0.266 & 0.319 & 0.297 & 0.234 & 0.194\tabularnewline
Adjusted R\textsuperscript{2}  & 0.170 & 0.208 & 0.264 & 0.241 & 0.178 & 0.135\tabularnewline
F Statistic & 4.002{*}{*}{*}  & 5.559{*}{*}{*} & 5.880{*}{*}{*} & 5.301{*}{*}{*} & 4.159{*}{*}{*} & 3.281{*}{*}{*}\tabularnewline
\hline\hline &  &  &  &  &  & \tabularnewline
\textit{Note:} & \multicolumn{6}{c}{$^{*}$p$<$0.1; $^{**}$p$<$0.05; $^{***}$p$<$0.01}\tabularnewline
\end{tabular}}
\par\end{centering}
\caption{Second-step regression of the estimates of rent gradients against
city characteristics.\label{tab:robustness_rent_second_step}}
\end{table}

\subsection{Robustness checks - Density gradients}

With our main specification (subsection \ref{subsec:Density-elasticities}),
we find density gradients of the expected direction in 192 cities
with OLS and 192 with 2SLS. We investigate whether this result holds
under the four robustness checks (Table \ref{tab:robustness_density_direction}).

\begin{table}[H]
\centering{}%
\begin{tabular}{|c|c|c|c|c|}
\hline 
 &  & Expected & Opposite & Non-significant\tabularnewline
 &  & direction & direction & (10\% level)\tabularnewline
\hline 
Main specification & OLS & 192 & 0 & 0\tabularnewline
Main specification & 2SLS & 192 & 0 & 0\tabularnewline
Robustness check 1 & OLS & 192 & 0 & 0\tabularnewline
Robustness check 2 & OLS & 192 & 0 & 0\tabularnewline
Robustness check 2 & 2SLS & 192 & 0 & 0\tabularnewline
Robustness check 3 & OLS & 192 & 0 & 0\tabularnewline
Robustness check 3 & 2SLS & 192 & 0 & 0\tabularnewline
Robustness check 4 & OLS & 192 & 0 & 0\tabularnewline
Robustness check 4 & 2SLS & 192 & 0 & 0\tabularnewline
\hline 
\end{tabular}\caption{Robustness checks -- direction of the density gradients.\label{tab:robustness_density_direction}}
\end{table}

With our main specification (subsection \ref{subsec:Density-elasticities}),
we find that the median R\textsuperscript{2} of regression \ref{eq:empirical_density}
is 0.17 with OLS and 0.15 with 2SLS (Table \ref{tab:r2_density}).
We investigate whether this result holds under the four robustness
checks (Table \ref{tab:robustness_density_r2}).

\begin{table}[H]
\centering{}%
\begin{tabular}{|c|c|c|c|c|c|c|c|}
\hline 
 &  & Mean & Min. & Q1 & Med. & Q3 & Max.\tabularnewline
\hline 
Main specification & OLS & 0.19 & 0.02 & 0.12 & 0.17 & 0.24 & 0.46\tabularnewline
Main specification & 2SLS & 0.15 & -1.12 & 0.09 & 0.15 & 0.23 & 0.46\tabularnewline
Robustness check 1 & OLS & 0.18 & 0.03 & 0.12 & 0.17 & 0.24 & 0.56\tabularnewline
Robustness check 2 & OLS & 0.23 & 0.05 & 0.16 & 0.22 & 0.29 & 0.61\tabularnewline
Robustness check 2 & 2SLS & 0.23 & 0.04 & 0.16 & 0.22 & 0.29 & 0.61\tabularnewline
Robustness check 3 & OLS & 0.21 & 0.03 & 0.15 & 0.20 & 0.26 & 0.54\tabularnewline
Robustness check 3 & 2SLS & 0.20 & -0.19 & 0.14 & 0.19 & 0.25 & 0.53\tabularnewline
Robustness check 4 & OLS & 0.19 & 0.02 & 0.12 & 0.18 & 0.24 & 0.46\tabularnewline
Robustness check 4 & 2SLS & 0.15 & -1.12 & 0.08 & 0.15 & 0.23 & 0.46\tabularnewline
\hline 
\end{tabular}\caption{Robustness checks - R\protect\textsuperscript{2} of equation \ref{eq:empirical_density}.\label{tab:robustness_density_r2}}
\end{table}

We then display the distribution of the density gradients estimated
in Robustness check 1. On average, moving 1km farther from the city
center reduces poplation density by 8.5\%.

\begin{figure}[H]
\centering

\subfloat[Histogram - distribution of parameter \textit{h'}\protect\textsubscript{\textit{c}}\textit{.}]{\centering\includegraphics[scale=0.7]{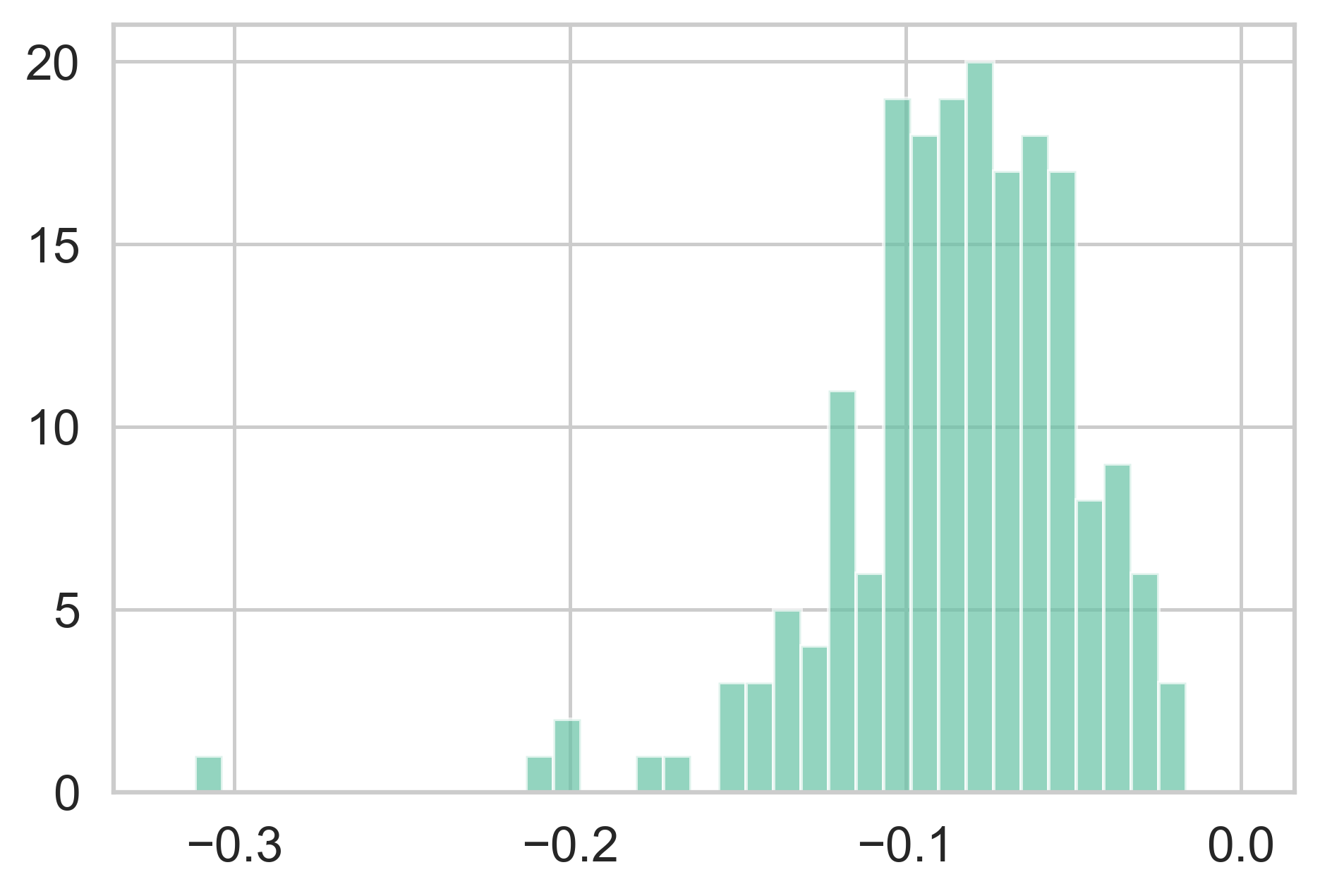}

}

\subfloat[Table - distribution of parameter \textit{h'}\protect\textsubscript{\textit{c}}\textit{.}]{\centering%
\begin{tabular}{cccccccc}
 & Mean & Min. & Q1 & Med. & Q3 & Max. & Nb. of obs.\tabularnewline
\hline 
OLS & -0.085 & -0.309 & -0.102 & -0.081 & -0.060 & -0.017 & 192\tabularnewline
\hline 
\end{tabular}

}\caption{Distribution of the estimates of parameter \textit{h'}\protect\textsubscript{\textit{c}}
(density gradient estimated with Robustness check 1) in the 192 cities.\label{fig:distrib_density_robustness}}
\end{figure}

With our main specification (subsection \ref{subsec:Density-elasticities}),
we find that the population, the monocentricity index, the percentage
of informal housing, the regulatory environment, and the dummies indicating
whether the city is located in Africa and Oceania explain the estimated
rent gradients (Table \ref{tab:second_step_density}). We investigate
whether this result holds under the four robustness checks (Table
\ref{tab:robustness_density_second_step}).

\begin{table}[H]
\begin{centering}
\begin{tabular}{cccccc}
\hline 
 & \multicolumn{5}{c}{\textit{Dependent variable: density gradients}}\tabularnewline
\cline{2-6} \cline{3-6} \cline{4-6} \cline{5-6} \cline{6-6} 
 & Main & Robustness & Robustness & Robustness & Robustness\tabularnewline
 & specification & check 1 & check 2 & check 3 & check 4\tabularnewline
\hline 
log(population) & -5.428{*}{*}{*}  & 0.018{*}{*}{*} & -0.155{*}{*} & -0.159{*} & -5.735{*}{*}{*}\tabularnewline
 & (0.856) & (0.003) & (0.075) & (0.091) & (0.856)\tabularnewline
log(income) &  & -0.001 & 0.158 & 0.392{*}{*}{*} & \tabularnewline
 &  & (0.005) & (0.110) & (0.138) & \tabularnewline
Monocentricity index & 4.346{*}{*}{*}  & -0.007 & -0.316{*}{*} & -0.393{*}{*} & 4.843{*}{*}{*}\tabularnewline
 & (1.591) & (0.007) & (0.128) & (0.178) & (1.611)\tabularnewline
Gini index & -0.104 & 0.001 & 0.004 & 0.005 & -0.107\tabularnewline
 & (0.147) & (0.001) & (0.011) & (0.015) & (0.146)\tabularnewline
Percentage of informal housing & -0.176{*}{*}  & -0.000 & 0.004 & 0.015 & -0.172{*}{*}\tabularnewline
 & (0.070) & (0.000) & (0.008) & (0.011) & (0.070)\tabularnewline
Coastal city & 1.583 & -0.012{*}{*} & -0.133 & -0.225 & 1.553\tabularnewline
 & (1.508) & (0.006) & (0.116) & (0.152) & (1.508)\tabularnewline
Regulatory environment & 3.119{*}  & -0.016{*}{*}{*} & -0.141 & -0.147 & 2.868\tabularnewline
 & (1.749) & (0.005) & (0.111) & (0.153) & (1.741)\tabularnewline
Asia & 2.602 & -0.015 & 0.932{*}{*}{*} & 1.015{*}{*}{*} & 2.963\tabularnewline
 & (3.939) & (0.014) & (0.270) & (0.377) & (3.963)\tabularnewline
Africa & -8.157{*}{*}{*}  & -0.002 & 1.067{*}{*}{*} & 1.094{*}{*} & -8.148{*}{*}{*}\tabularnewline
 & (2.968) & (0.014) & (0.280) & (0.452) & (3.953)\tabularnewline
Oceania & -9.958{*}{*}{*}  & 0.034{*}{*}{*} & 0.059 & 0.138 & -10.992{*}{*}{*}\tabularnewline
 & (2.442) & (0.008) & (0.324) & (0.397) & (2.283)\tabularnewline
North America & 2.598 & 0.022{*}{*}{*} & 0.379 & 0.627{*}{*} & 2.682\tabularnewline
 & (3.429) & (0.010) & (0.238) & (0.296) & (3.423)\tabularnewline
South America & -3.264 & 0.022{*} & 0.228 & -0.038 & -3.503\tabularnewline
 & (3.801) & (0.016) & (0.264) & (0.402) & (3.814)\tabularnewline
Constant & 95.918{*}{*}{*}  & -0.329{*}{*}{*} & -1.772 & -4.628{*}{*} & 99.449{*}{*}{*}\tabularnewline
 & (12.958) & (0.076) & (1.677) & (2.026) & (13.071)\tabularnewline
\hline &  &  &  &  & \tabularnewline
Observations & 192 & 192 & 192 & 192 & 192\tabularnewline
R\textsuperscript{2} & 0.508 & 0.319 & 0.213 & 0.184 & 0.525\tabularnewline
Adjusted R\textsuperscript{2 } & 0.478 & 0.273 & 0.161 & 0.129 & 0.496\tabularnewline
F Statistic & 29.49{*}{*}{*}  & 10.44{*}{*}{*} & 4.990{*}{*}{*} & 4.212{*}{*}{*} & 29.85{*}{*}{*}\tabularnewline
\hline\hline &  &  &  &  & \tabularnewline
\textit{Note:} & \multicolumn{5}{c}{$^{*}$p$<$0.1; $^{**}$p$<$0.05; $^{***}$p$<$0.01}\tabularnewline
\end{tabular}\caption{Second-step regression of the estimates of density gradients against
city characteristics. \label{tab:robustness_density_second_step}}
\par\end{centering}
\end{table}

\section{Regulatory environmnent\label{appendix:Regulatory-environmnent}}

When the categorical variable coding for the regulatory environment
is different from 0 based on the authors' judgements, an explanation
is provided as a footnote.

\begin{longtable}[c]{|c|c|c|c|}
\hline 
{\footnotesize{}City} & {\footnotesize{}Country} & {\footnotesize{}Regulatory env.} & {\footnotesize{}Source}\tabularnewline
\hline 
\hline
\endhead
{\footnotesize{}Abidjan} & {\footnotesize{}Côte d'Ivoire} & {\footnotesize{}1} & {\footnotesize{}Bertaud and Malpezzi (2014)}\tabularnewline
{\footnotesize{}Adana} & {\footnotesize{}Turkey} & {\footnotesize{}0} & {\footnotesize{}Authors' judgement}\tabularnewline
{\footnotesize{}Addis\_Ababa} & {\footnotesize{}Ethiopia} & {\footnotesize{}0} & {\footnotesize{}Authors' judgement}\tabularnewline
{\footnotesize{}Adelaide} & {\footnotesize{}Australia} & {\footnotesize{}1} & {\footnotesize{}Authors' judgement}\footnote{{\footnotesize{}Adelaide is a planned city, designed in 1836 by the
first surveyor-general of South Australia, Colonel William Light.
Adelaide was designed as a grid, with five squares in the city center,
surrounded by a ring of parks.}}\tabularnewline
{\footnotesize{}Ahmedabad} & {\footnotesize{}India} & {\footnotesize{}1} & {\footnotesize{}Bertaud and Malpezzi (2014)}\tabularnewline
{\footnotesize{}Amsterdam} & {\footnotesize{}Netherlands} & {\footnotesize{}0} & {\footnotesize{}Authors' judgement}\tabularnewline
{\footnotesize{}Ankara} & {\footnotesize{}Turkey} & {\footnotesize{}0} & {\footnotesize{}Authors' judgement}\tabularnewline
{\footnotesize{}Antwerp} & {\footnotesize{}Belgium} & {\footnotesize{}0} & {\footnotesize{}Authors' judgement}\tabularnewline
{\footnotesize{}Arequipa} & {\footnotesize{}Peru} & {\footnotesize{}0} & {\footnotesize{}Authors' judgement}\tabularnewline
{\footnotesize{}Athens} & {\footnotesize{}Greece} & {\footnotesize{}0} & {\footnotesize{}Authors' judgement}\tabularnewline
{\footnotesize{}Atlanta} & {\footnotesize{}United States of America} & {\footnotesize{}0} & {\footnotesize{}Bertaud and Malpezzi (2014)}\tabularnewline
{\footnotesize{}Auckland} & {\footnotesize{}New Zealand} & {\footnotesize{}0} & {\footnotesize{}Authors' judgement}\tabularnewline
{\footnotesize{}Bandung} & {\footnotesize{}Indonesia} & {\footnotesize{}0} & {\footnotesize{}Authors' judgement}\tabularnewline
{\footnotesize{}Bangalore} & {\footnotesize{}India} & {\footnotesize{}1} & {\footnotesize{}Bertaud and Malpezzi (2014)}\tabularnewline
{\footnotesize{}Bangkok} & {\footnotesize{}Thailand} & {\footnotesize{}0} & {\footnotesize{}Bertaud and Malpezzi (2014)}\tabularnewline
{\footnotesize{}Barcelona} & {\footnotesize{}Spain} & {\footnotesize{}0} & {\footnotesize{}Bertaud and Malpezzi (2014)}\tabularnewline
{\footnotesize{}Barranquilla} & {\footnotesize{}Colombia} & {\footnotesize{}0} & {\footnotesize{}Authors' judgement}\tabularnewline
{\footnotesize{}Basel} & {\footnotesize{}Switzerland} & {\footnotesize{}0} & {\footnotesize{}Authors' judgement}\tabularnewline
{\footnotesize{}Beijing} & {\footnotesize{}China} & {\footnotesize{}0} & {\footnotesize{}Bertaud and Malpezzi (2014)}\tabularnewline
{\footnotesize{}Belem} & {\footnotesize{}Brazil} & {\footnotesize{}0} & {\footnotesize{}Authors' judgement}\tabularnewline
{\footnotesize{}Belo\_Horizonte} & {\footnotesize{}Brazil} & {\footnotesize{}1} & {\footnotesize{}Authors' judgement}\footnote{{\footnotesize{}Belo Horizonte was created as the new state capital
in the 1890s. It was Brazil’s first planned city. In particular, its
urban core has a very specific street pattern, with a wide grid of
broad ceremonial avenues and a regular street grid, offset at a 45-degree
angle, and a city square at the intersections of the streets of the
two grids.}}\tabularnewline
{\footnotesize{}Bergen} & {\footnotesize{}Norway} & {\footnotesize{}0} & {\footnotesize{}Authors' judgement}\tabularnewline
{\footnotesize{}Berlin} & {\footnotesize{}Germany} & {\footnotesize{}1} & {\footnotesize{}Bertaud and Malpezzi (2014)}\tabularnewline
{\footnotesize{}Bern} & {\footnotesize{}Switzerland} & {\footnotesize{}0} & {\footnotesize{}Authors' judgement}\tabularnewline
{\footnotesize{}Bilbao} & {\footnotesize{}Spain} & {\footnotesize{}0} & {\footnotesize{}Authors' judgement}\tabularnewline
{\footnotesize{}Birmingham} & {\footnotesize{}United Kingdom} & {\footnotesize{}0} & {\footnotesize{}Authors' judgement}\tabularnewline
{\footnotesize{}Bogota} & {\footnotesize{}Colombia} & {\footnotesize{}0} & {\footnotesize{}Authors' judgement}\tabularnewline
{\footnotesize{}Bordeaux} & {\footnotesize{}France} & {\footnotesize{}0} & {\footnotesize{}Authors' judgement}\tabularnewline
{\footnotesize{}Braga} & {\footnotesize{}Portugal} & {\footnotesize{}0} & {\footnotesize{}Authors' judgement}\tabularnewline
{\footnotesize{}Brasilia} & {\footnotesize{}Brazil} & {\footnotesize{}2} & {\footnotesize{}Bertaud and Malpezzi (2014)}\tabularnewline
{\footnotesize{}Bremen} & {\footnotesize{}Germany} & {\footnotesize{}0} & {\footnotesize{}Authors' judgement}\tabularnewline
{\footnotesize{}Brisbane} & {\footnotesize{}Australia} & {\footnotesize{}0} & {\footnotesize{}Authors' judgement}\tabularnewline
{\footnotesize{}Bristol} & {\footnotesize{}United Kingdom} & {\footnotesize{}0} & {\footnotesize{}Authors' judgement}\tabularnewline
{\footnotesize{}Brussels} & {\footnotesize{}Belgium} & {\footnotesize{}0} & {\footnotesize{}Authors' judgement}\tabularnewline
{\footnotesize{}Bucharest} & {\footnotesize{}Romania} & {\footnotesize{}0} & {\footnotesize{}Authors' judgement}\tabularnewline
{\footnotesize{}Budapest} & {\footnotesize{}Hungary} & {\footnotesize{}1} & {\footnotesize{}Bertaud and Malpezzi (2014)}\tabularnewline
{\footnotesize{}Buenos\_Aires} & {\footnotesize{}Argentina} & {\footnotesize{}0} & {\footnotesize{}Bertaud and Malpezzi (2014)}\tabularnewline
{\footnotesize{}Bursa} & {\footnotesize{}Turkey} & {\footnotesize{}0} & {\footnotesize{}Authors' judgement}\tabularnewline
{\footnotesize{}Cali} & {\footnotesize{}Colombia} & {\footnotesize{}0} & {\footnotesize{}Authors' judgement}\tabularnewline
{\footnotesize{}Cape\_Town} & {\footnotesize{}South Africa} & {\footnotesize{}2} & {\footnotesize{}Bertaud and Malpezzi (2014)}\tabularnewline
{\footnotesize{}Cartagena} & {\footnotesize{}Colombia} & {\footnotesize{}0} & {\footnotesize{}Authors' judgement}\tabularnewline
{\footnotesize{}Casablanca} & {\footnotesize{}Morocco} & {\footnotesize{}0} & {\footnotesize{}Authors' judgement}\tabularnewline
{\footnotesize{}Charleroi} & {\footnotesize{}Belgium} & {\footnotesize{}0} & {\footnotesize{}Authors' judgement}\tabularnewline
{\footnotesize{}Chelyabinsk} & {\footnotesize{}Russian Federation} & {\footnotesize{}0} & {\footnotesize{}Authors' judgement}\tabularnewline
{\footnotesize{}Chiang\_Mai} & {\footnotesize{}Thailand} & {\footnotesize{}0} & {\footnotesize{}Authors' judgement}\tabularnewline
{\footnotesize{}Chicago} & {\footnotesize{}United States of America} & {\footnotesize{}0} & {\footnotesize{}Bertaud and Malpezzi (2014)}\tabularnewline
{\footnotesize{}Christchurch} & {\footnotesize{}New Zealand} & {\footnotesize{}0} & {\footnotesize{}Authors' judgement}\tabularnewline
{\footnotesize{}Cluj\_Napoca} & {\footnotesize{}Romania} & {\footnotesize{}0} & {\footnotesize{}Authors' judgement}\tabularnewline
{\footnotesize{}Coimbra} & {\footnotesize{}Portugal} & {\footnotesize{}0} & {\footnotesize{}Authors' judgement}\tabularnewline
{\footnotesize{}Concepcion} & {\footnotesize{}Chile} & {\footnotesize{}0} & {\footnotesize{}Authors' judgement}\tabularnewline
{\footnotesize{}Cordoba} & {\footnotesize{}Argentina} & {\footnotesize{}0} & {\footnotesize{}Authors' judgement}\tabularnewline
{\footnotesize{}Cork} & {\footnotesize{}Ireland} & {\footnotesize{}0} & {\footnotesize{}Authors' judgement}\tabularnewline
{\footnotesize{}Cracow} & {\footnotesize{}Poland} & {\footnotesize{}1} & {\footnotesize{}Bertaud and Malpezzi (2014)}\tabularnewline
{\footnotesize{}Curitiba} & {\footnotesize{}Brazil} & {\footnotesize{}2} & {\footnotesize{}Bertaud and Malpezzi (2014)}\tabularnewline
{\footnotesize{}Delhi} & {\footnotesize{}India} & {\footnotesize{}1} & {\footnotesize{}Authors' judgement}\footnote{{\footnotesize{}Delhi has a long history of urban planning. The street
network was designed for defense purposes, with some transverse streets
leading from one major gate to another. In 1911, the decision was
made to transfer the capital to Delhi, and the new city of New Delhi
was planned inside the existing Delhi. New Delhi was planned with
wide, straight streets, trees in double rows on either side and a
broad central avenue. The Delhi Development Authority’s 20-year (1962-1981)
plan, the first Master Plan for the whole of Delhi, also largely impacted
land use in the city.}}\tabularnewline
{\footnotesize{}Detroit} & {\footnotesize{}United States of America} & {\footnotesize{}0} & {\footnotesize{}Authors' judgement}\tabularnewline
{\footnotesize{}Dresden} & {\footnotesize{}Germany} & {\footnotesize{}0} & {\footnotesize{}Authors' judgement}\tabularnewline
{\footnotesize{}Dublin} & {\footnotesize{}Ireland} & {\footnotesize{}0} & {\footnotesize{}Authors' judgement}\tabularnewline
{\footnotesize{}Edinburgh} & {\footnotesize{}United Kingdom} & {\footnotesize{}0} & {\footnotesize{}Authors' judgement}\tabularnewline
{\footnotesize{}Faisalabad} & {\footnotesize{}Pakistan} & {\footnotesize{}0} & {\footnotesize{}Authors' judgement}\tabularnewline
{\footnotesize{}Fez} & {\footnotesize{}Morocco} & {\footnotesize{}0} & {\footnotesize{}Authors' judgement}\tabularnewline
{\footnotesize{}Fortaleza} & {\footnotesize{}Brazil} & {\footnotesize{}0} & {\footnotesize{}Authors' judgement}\tabularnewline
{\footnotesize{}Frankfurt\_am\_Main} & {\footnotesize{}Germany} & {\footnotesize{}0} & {\footnotesize{}Authors' judgement}\tabularnewline
{\footnotesize{}Geneva} & {\footnotesize{}Switzerland} & {\footnotesize{}0} & {\footnotesize{}Authors' judgement}\tabularnewline
{\footnotesize{}Genoa} & {\footnotesize{}Italy} & {\footnotesize{}0} & {\footnotesize{}Authors' judgement}\tabularnewline
{\footnotesize{}Ghent} & {\footnotesize{}Belgium} & {\footnotesize{}0} & {\footnotesize{}Authors' judgement}\tabularnewline
{\footnotesize{}Glasgow} & {\footnotesize{}United Kingdom} & {\footnotesize{}0} & {\footnotesize{}Authors' judgement}\tabularnewline
{\footnotesize{}Goiania} & {\footnotesize{}Brazil} & {\footnotesize{}0} & {\footnotesize{}Authors' judgement}\tabularnewline
{\footnotesize{}Gothenburg} & {\footnotesize{}Sweden} & {\footnotesize{}0} & {\footnotesize{}Authors' judgement}\tabularnewline
{\footnotesize{}Guadalajara} & {\footnotesize{}Mexico} & {\footnotesize{}0} & {\footnotesize{}Authors' judgement}\tabularnewline
{\footnotesize{}Hamburg} & {\footnotesize{}Germany} & {\footnotesize{}0} & {\footnotesize{}Authors' judgement}\tabularnewline
{\footnotesize{}Hanover} & {\footnotesize{}Germany} & {\footnotesize{}0} & {\footnotesize{}Authors' judgement}\tabularnewline
{\footnotesize{}Helsinki} & {\footnotesize{}Finland} & {\footnotesize{}0} & {\footnotesize{}Authors' judgement}\tabularnewline
{\footnotesize{}Hong\_Kong} & {\footnotesize{}China, Hong Kong SAR} & {\footnotesize{}0} & {\footnotesize{}Bertaud and Malpezzi (2014)}\tabularnewline
{\footnotesize{}Houston} & {\footnotesize{}United States of America} & {\footnotesize{}0} & {\footnotesize{}Bertaud and Malpezzi (2014)}\tabularnewline
{\footnotesize{}Hyderabad} & {\footnotesize{}India} & {\footnotesize{}1} & {\footnotesize{}Bertaud and Malpezzi (2014)}\tabularnewline
{\footnotesize{}Isfahan} & {\footnotesize{}Iran (Islamic Republic of)} & {\footnotesize{}0} & {\footnotesize{}Authors' judgement}\tabularnewline
{\footnotesize{}Istanbul} & {\footnotesize{}Turkey} & {\footnotesize{}1} & {\footnotesize{}Authors' judgement}\footnote{{\footnotesize{}Istanbul has a long history of urban planning. Major
developments occurred before World War II, with the idea of comprehensive
city development in Istanbul introduced in plans drawn up by city
planner Henri Prost and in a planning competition organized in 1933.
In the second half of the nineteenth century, comprehensive planning
on a metropolitan scale was institutionalized.}}\tabularnewline
{\footnotesize{}Izmir} & {\footnotesize{}Turkey} & {\footnotesize{}0} & {\footnotesize{}Authors' judgement}\tabularnewline
{\footnotesize{}Jakarta} & {\footnotesize{}Indonesia} & {\footnotesize{}0} & {\footnotesize{}Bertaud and Malpezzi (2014)}\tabularnewline
{\footnotesize{}Jinan} & {\footnotesize{}China} & {\footnotesize{}0} & {\footnotesize{}Authors' judgement}\tabularnewline
{\footnotesize{}Johannesburg} & {\footnotesize{}South Africa} & {\footnotesize{}2} & {\footnotesize{}Bertaud and Malpezzi (2014)}\tabularnewline
{\footnotesize{}Karachi} & {\footnotesize{}Pakistan} & {\footnotesize{}0} & {\footnotesize{}Authors' judgement}\tabularnewline
{\footnotesize{}Kazan} & {\footnotesize{}Russian Federation} & {\footnotesize{}0} & {\footnotesize{}Authors' judgement}\tabularnewline
{\footnotesize{}Lahore} & {\footnotesize{}Pakistan} & {\footnotesize{}0} & {\footnotesize{}Authors' judgement}\tabularnewline
{\footnotesize{}Lausanne} & {\footnotesize{}Switzerland} & {\footnotesize{}0} & {\footnotesize{}Authors' judgement}\tabularnewline
{\footnotesize{}Leeds} & {\footnotesize{}United Kingdom} & {\footnotesize{}0} & {\footnotesize{}Authors' judgement}\tabularnewline
{\footnotesize{}Leicester} & {\footnotesize{}United Kingdom} & {\footnotesize{}0} & {\footnotesize{}Authors' judgement}\tabularnewline
{\footnotesize{}Leipzig} & {\footnotesize{}Germany} & {\footnotesize{}0} & {\footnotesize{}Authors' judgement}\tabularnewline
{\footnotesize{}Lille} & {\footnotesize{}France} & {\footnotesize{}0} & {\footnotesize{}Authors' judgement}\tabularnewline
{\footnotesize{}Lima} & {\footnotesize{}Peru} & {\footnotesize{}0} & {\footnotesize{}Authors' judgement}\tabularnewline
{\footnotesize{}Lisbon} & {\footnotesize{}Portugal} & {\footnotesize{}0} & {\footnotesize{}Authors' judgement}\tabularnewline
{\footnotesize{}Liverpool} & {\footnotesize{}United Kingdom} & {\footnotesize{}0} & {\footnotesize{}Authors' judgement}\tabularnewline
{\footnotesize{}Ljubljana} & {\footnotesize{}Slovenia} & {\footnotesize{}1} & {\footnotesize{}Bertaud and Malpezzi (2014)}\tabularnewline
{\footnotesize{}Lodz} & {\footnotesize{}Poland} & {\footnotesize{}0} & {\footnotesize{}Authors' judgement}\tabularnewline
{\footnotesize{}London} & {\footnotesize{}United Kingdom} & {\footnotesize{}0} & {\footnotesize{}Bertaud and Malpezzi (2014)}\tabularnewline
{\footnotesize{}Los\_Angeles} & {\footnotesize{}United States of America} & {\footnotesize{}0} & {\footnotesize{}Bertaud and Malpezzi (2014)}\tabularnewline
{\footnotesize{}Lyon} & {\footnotesize{}France} & {\footnotesize{}0} & {\footnotesize{}Authors' judgement}\tabularnewline
{\footnotesize{}Madrid} & {\footnotesize{}Spain} & {\footnotesize{}0} & {\footnotesize{}Authors' judgement}\tabularnewline
{\footnotesize{}Malaga} & {\footnotesize{}Spain} & {\footnotesize{}0} & {\footnotesize{}Authors' judgement}\tabularnewline
{\footnotesize{}Malmo} & {\footnotesize{}Sweden} & {\footnotesize{}0} & {\footnotesize{}Authors' judgement}\tabularnewline
{\footnotesize{}Manaus} & {\footnotesize{}Brazil} & {\footnotesize{}0} & {\footnotesize{}Authors' judgement}\tabularnewline
{\footnotesize{}Marseille} & {\footnotesize{}France} & {\footnotesize{}0} & {\footnotesize{}Bertaud and Malpezzi (2014)}\tabularnewline
{\footnotesize{}Mar\_del\_Plata} & {\footnotesize{}Argentina} & {\footnotesize{}0} & {\footnotesize{}Authors' judgement}\tabularnewline
{\footnotesize{}Mashhad} & {\footnotesize{}Iran (Islamic Republic of)} & {\footnotesize{}0} & {\footnotesize{}Authors' judgement}\tabularnewline
{\footnotesize{}Medan} & {\footnotesize{}Indonesia} & {\footnotesize{}0} & {\footnotesize{}Authors' judgement}\tabularnewline
{\footnotesize{}Medellin} & {\footnotesize{}Colombia} & {\footnotesize{}0} & {\footnotesize{}Authors' judgement}\tabularnewline
{\footnotesize{}Melbourne} & {\footnotesize{}Australia} & {\footnotesize{}0} & {\footnotesize{}Authors' judgement}\tabularnewline
{\footnotesize{}Mexico\_City} & {\footnotesize{}Mexico} & {\footnotesize{}0} & {\footnotesize{}Bertaud and Malpezzi (2014)}\tabularnewline
{\footnotesize{}Milan} & {\footnotesize{}Italy} & {\footnotesize{}0} & {\footnotesize{}Authors' judgement}\tabularnewline
{\footnotesize{}Monterrey} & {\footnotesize{}Mexico} & {\footnotesize{}0} & {\footnotesize{}Authors' judgement}\tabularnewline
{\footnotesize{}Montevideo} & {\footnotesize{}Uruguay} & {\footnotesize{}0} & {\footnotesize{}Authors' judgement}\tabularnewline
{\footnotesize{}Montpellier} & {\footnotesize{}France} & {\footnotesize{}0} & {\footnotesize{}Authors' judgement}\tabularnewline
{\footnotesize{}Moscow} & {\footnotesize{}Russian Federation} & {\footnotesize{}2} & {\footnotesize{}Bertaud and Malpezzi (2014)}\tabularnewline
{\footnotesize{}Mumbai} & {\footnotesize{}India} & {\footnotesize{}1} & {\footnotesize{}Bertaud and Malpezzi (2014)}\tabularnewline
{\footnotesize{}Munich} & {\footnotesize{}Germany} & {\footnotesize{}0} & {\footnotesize{}Authors' judgement}\tabularnewline
{\footnotesize{}Nantes} & {\footnotesize{}France} & {\footnotesize{}0} & {\footnotesize{}Authors' judgement}\tabularnewline
{\footnotesize{}Naples} & {\footnotesize{}Italy} & {\footnotesize{}0} & {\footnotesize{}Authors' judgement}\tabularnewline
{\footnotesize{}New\_York} & {\footnotesize{}United States of America} & {\footnotesize{}0} & {\footnotesize{}Bertaud and Malpezzi (2014)}\tabularnewline
{\footnotesize{}Nice} & {\footnotesize{}France} & {\footnotesize{}0} & {\footnotesize{}Authors' judgement}\tabularnewline
{\footnotesize{}Nizhny\_Novgorod} & {\footnotesize{}Russian Federation} & {\footnotesize{}0} & {\footnotesize{}Authors' judgement}\tabularnewline
{\footnotesize{}Nottingham} & {\footnotesize{}United Kingdom} & {\footnotesize{}0} & {\footnotesize{}Authors' judgement}\tabularnewline
{\footnotesize{}Novosibirsk} & {\footnotesize{}Russian Federation} & {\footnotesize{}0} & {\footnotesize{}Authors' judgement}\tabularnewline
{\footnotesize{}Nuremberg} & {\footnotesize{}Germany} & {\footnotesize{}0} & {\footnotesize{}Authors' judgement}\tabularnewline
{\footnotesize{}Omsk} & {\footnotesize{}Russian Federation} & {\footnotesize{}0} & {\footnotesize{}Authors' judgement}\tabularnewline
{\footnotesize{}Oslo} & {\footnotesize{}Norway} & {\footnotesize{}0} & {\footnotesize{}Authors' judgement}\tabularnewline
{\footnotesize{}Palermo} & {\footnotesize{}Italy} & {\footnotesize{}0} & {\footnotesize{}Authors' judgement}\tabularnewline
{\footnotesize{}Paris} & {\footnotesize{}France} & {\footnotesize{}0} & {\footnotesize{}Bertaud and Malpezzi (2014)}\tabularnewline
{\footnotesize{}Patras} & {\footnotesize{}Greece} & {\footnotesize{}0} & {\footnotesize{}Authors' judgement}\tabularnewline
{\footnotesize{}Perth} & {\footnotesize{}Australia} & {\footnotesize{}0} & {\footnotesize{}Authors' judgement}\tabularnewline
{\footnotesize{}Portland} & {\footnotesize{}United States of America} & {\footnotesize{}1} & {\footnotesize{}Bertaud and Malpezzi (2014)}\tabularnewline
{\footnotesize{}Porto} & {\footnotesize{}Portugal} & {\footnotesize{}0} & {\footnotesize{}Authors' judgement}\tabularnewline
{\footnotesize{}Porto\_Alegre} & {\footnotesize{}Brazil} & {\footnotesize{}0} & {\footnotesize{}Authors' judgement}\tabularnewline
{\footnotesize{}Poznan} & {\footnotesize{}Poland} & {\footnotesize{}0} & {\footnotesize{}Authors' judgement}\tabularnewline
{\footnotesize{}Prague} & {\footnotesize{}Czechia} & {\footnotesize{}1} & {\footnotesize{}Bertaud and Malpezzi (2014)}\tabularnewline
{\footnotesize{}Puebla} & {\footnotesize{}Mexico} & {\footnotesize{}1} & {\footnotesize{}Authors' judgement}\footnote{{\footnotesize{}Puebla was the first planned city in the Americas
with streets traced in rectangular 100m by 200m blocks, and where
the city's whole central district is oriented to the movement of the
sun in the sky. Most of the major arteries follow the regular orientation
of the plan, but one main street, the “Diagonale de Los Defensores
de la Republica,” crosses the city on a diagonal. Puebla’s Zocalo,
the central esplanade of the historic area, is considered one of the
great achievements of colonial urban planning.}}\tabularnewline
{\footnotesize{}Rabat} & {\footnotesize{}Morocco} & {\footnotesize{}0} & {\footnotesize{}Authors' judgement}\tabularnewline
{\footnotesize{}Recife} & {\footnotesize{}Brazil} & {\footnotesize{}0} & {\footnotesize{}Authors' judgement}\tabularnewline
{\footnotesize{}Rennes} & {\footnotesize{}France} & {\footnotesize{}0} & {\footnotesize{}Authors' judgement}\tabularnewline
{\footnotesize{}Riga} & {\footnotesize{}Latvia} & {\footnotesize{}1} & {\footnotesize{}Bertaud and Malpezzi (2014)}\tabularnewline
{\footnotesize{}Rio\_de\_Janeiro} & {\footnotesize{}Brazil} & {\footnotesize{}0} & {\footnotesize{}Bertaud and Malpezzi (2014)}\tabularnewline
{\footnotesize{}Rome} & {\footnotesize{}Italy} & {\footnotesize{}0} & {\footnotesize{}Authors' judgement}\tabularnewline
{\footnotesize{}Rosario} & {\footnotesize{}Argentina} & {\footnotesize{}0} & {\footnotesize{}Authors' judgement}\tabularnewline
{\footnotesize{}Rostov\_on\_Don} & {\footnotesize{}Russian Federation} & {\footnotesize{}0} & {\footnotesize{}Authors' judgement}\tabularnewline
{\footnotesize{}Rotterdam} & {\footnotesize{}Netherlands} & {\footnotesize{}0} & {\footnotesize{}Authors' judgement}\tabularnewline
{\footnotesize{}Salvador} & {\footnotesize{}Brazil} & {\footnotesize{}0} & {\footnotesize{}Authors' judgement}\tabularnewline
{\footnotesize{}Samara} & {\footnotesize{}Russian Federation} & {\footnotesize{}0} & {\footnotesize{}Authors' judgement}\tabularnewline
{\footnotesize{}Santiago} & {\footnotesize{}Chile} & {\footnotesize{}0} & {\footnotesize{}Authors' judgement}\tabularnewline
{\footnotesize{}San\_Diego} & {\footnotesize{}United States of America} & {\footnotesize{}0} & {\footnotesize{}Authors' judgement}\tabularnewline
{\footnotesize{}San\_Fransisco} & {\footnotesize{}United States of America} & {\footnotesize{}1} & {\footnotesize{}Bertaud and Malpezzi (2014)}\tabularnewline
{\footnotesize{}Sao\_Paulo} & {\footnotesize{}Brazil} & {\footnotesize{}0} & {\footnotesize{}Authors' judgement}\tabularnewline
{\footnotesize{}Seattle} & {\footnotesize{}United States of America} & {\footnotesize{}0} & {\footnotesize{}Authors' judgement}\tabularnewline
{\footnotesize{}Seville} & {\footnotesize{}Spain} & {\footnotesize{}0} & {\footnotesize{}Authors' judgement}\tabularnewline
{\footnotesize{}Sfax} & {\footnotesize{}Tunisia} & {\footnotesize{}0} & {\footnotesize{}Authors' judgement}\tabularnewline
{\footnotesize{}Shanghai} & {\footnotesize{}China} & {\footnotesize{}0} & {\footnotesize{}Bertaud and Malpezzi (2014)}\tabularnewline
{\footnotesize{}Singapore} & {\footnotesize{}Singapore} & {\footnotesize{}0} & {\footnotesize{}Bertaud and Malpezzi (2014)}\tabularnewline
{\footnotesize{}Sofia} & {\footnotesize{}Bulgaria} & {\footnotesize{}1} & {\footnotesize{}Bertaud and Malpezzi (2014)}\tabularnewline
{\footnotesize{}Sousse} & {\footnotesize{}Tunisia} & {\footnotesize{}0} & {\footnotesize{}Authors' judgement}\tabularnewline
{\footnotesize{}Split} & {\footnotesize{}Croatia} & {\footnotesize{}0} & {\footnotesize{}Authors' judgement}\tabularnewline
{\footnotesize{}Stavanger} & {\footnotesize{}Norway} & {\footnotesize{}0} & {\footnotesize{}Authors' judgement}\tabularnewline
{\footnotesize{}Stockholm} & {\footnotesize{}Sweden} & {\footnotesize{}2} & {\footnotesize{}Bertaud and Malpezzi (2014)}\tabularnewline
{\footnotesize{}Strasbourg} & {\footnotesize{}France} & {\footnotesize{}0} & {\footnotesize{}Authors' judgement}\tabularnewline
{\footnotesize{}Stuttgart} & {\footnotesize{}Germany} & {\footnotesize{}0} & {\footnotesize{}Authors' judgement}\tabularnewline
{\footnotesize{}St\_Petersburg} & {\footnotesize{}Russian Federation} & {\footnotesize{}1} & {\footnotesize{}Bertaud and Malpezzi (2014)}\tabularnewline
{\footnotesize{}Surabaya} & {\footnotesize{}Indonesia} & {\footnotesize{}0} & {\footnotesize{}Authors' judgement}\tabularnewline
{\footnotesize{}Sydney} & {\footnotesize{}Australia} & {\footnotesize{}0} & {\footnotesize{}Authors' judgement}\tabularnewline
{\footnotesize{}Tampere} & {\footnotesize{}Finland} & {\footnotesize{}0} & {\footnotesize{}Authors' judgement}\tabularnewline
{\footnotesize{}Tehran} & {\footnotesize{}Iran (Islamic Republic of)} & {\footnotesize{}0} & {\footnotesize{}Authors' judgement}\tabularnewline
{\footnotesize{}Thessaloniki} & {\footnotesize{}Greece} & {\footnotesize{}0} & {\footnotesize{}Authors' judgement}\tabularnewline
{\footnotesize{}The\_Hague} & {\footnotesize{}Netherlands} & {\footnotesize{}0} & {\footnotesize{}Authors' judgement}\tabularnewline
{\footnotesize{}Tianjin} & {\footnotesize{}China} & {\footnotesize{}0} & {\footnotesize{}Bertaud and Malpezzi (2014)}\tabularnewline
{\footnotesize{}Tijuana} & {\footnotesize{}Mexico} & {\footnotesize{}0} & {\footnotesize{}Authors' judgement}\tabularnewline
{\footnotesize{}Timisoara} & {\footnotesize{}Romania} & {\footnotesize{}0} & {\footnotesize{}Authors' judgement}\tabularnewline
{\footnotesize{}Toluca} & {\footnotesize{}Mexico} & {\footnotesize{}0} & {\footnotesize{}Authors' judgement}\tabularnewline
{\footnotesize{}Toulouse} & {\footnotesize{}France} & {\footnotesize{}0} & {\footnotesize{}Bertaud and Malpezzi (2014)}\tabularnewline
{\footnotesize{}Trujillo} & {\footnotesize{}Peru} & {\footnotesize{}0} & {\footnotesize{}Authors' judgement}\tabularnewline
{\footnotesize{}Tunis} & {\footnotesize{}Tunisia} & {\footnotesize{}2} & {\footnotesize{}Bertaud and Malpezzi (2014)}\tabularnewline
{\footnotesize{}Turin} & {\footnotesize{}Italy} & {\footnotesize{}0} & {\footnotesize{}Authors' judgement}\tabularnewline
{\footnotesize{}Turku} & {\footnotesize{}Finland} & {\footnotesize{}0} & {\footnotesize{}Authors' judgement}\tabularnewline
{\footnotesize{}Ulan\_Bator} & {\footnotesize{}Mongolia} & {\footnotesize{}0} & {\footnotesize{}Authors' judgement}\tabularnewline
{\footnotesize{}Valencia} & {\footnotesize{}Spain} & {\footnotesize{}0} & {\footnotesize{}Authors' judgement}\tabularnewline
{\footnotesize{}Valparaiso} & {\footnotesize{}Chile} & {\footnotesize{}0} & {\footnotesize{}Authors' judgement}\tabularnewline
{\footnotesize{}Warsaw} & {\footnotesize{}Poland} & {\footnotesize{}1} & {\footnotesize{}Bertaud and Malpezzi (2014)}\tabularnewline
{\footnotesize{}Washington\_DC} & {\footnotesize{}United States of America} & {\footnotesize{}0} & {\footnotesize{}Bertaud and Malpezzi (2014)}\tabularnewline
{\footnotesize{}Wellington} & {\footnotesize{}New Zealand} & {\footnotesize{}0} & {\footnotesize{}Authors' judgement}\tabularnewline
{\footnotesize{}Wroclaw} & {\footnotesize{}Poland} & {\footnotesize{}0} & {\footnotesize{}Authors' judgement}\tabularnewline
{\footnotesize{}Wuhan} & {\footnotesize{}China} & {\footnotesize{}0} & {\footnotesize{}Authors' judgement}\tabularnewline
{\footnotesize{}Yekaterinburg} & {\footnotesize{}Russian Federation} & {\footnotesize{}0} & {\footnotesize{}Authors' judgement}\tabularnewline
{\footnotesize{}Yerevan} & {\footnotesize{}Armenia} & {\footnotesize{}2} & {\footnotesize{}Bertaud and Malpezzi (2014)}\tabularnewline
{\footnotesize{}Zagreb} & {\footnotesize{}Croatia} & {\footnotesize{}0} & {\footnotesize{}Authors' judgement}\tabularnewline
{\footnotesize{}Zhengzhou} & {\footnotesize{}China} & {\footnotesize{}0} & {\footnotesize{}Authors' judgement}\tabularnewline
{\footnotesize{}Zurich} & {\footnotesize{}Switzerland} & {\footnotesize{}0} & {\footnotesize{}Authors' judgement}\tabularnewline
\hline 
\end{longtable}
\end{document}